\documentclass[prb,twocolumn,aps,10pt,superscriptaddress]{revtex4}
\usepackage{amsmath}
\usepackage{graphicx}
\usepackage{subfigure}
\usepackage{hyperref}  % for PDF only
\usepackage{float}
\usepackage{color}

\begin{document}

\title{Oxygen vacancies and hydrogen doping in 
       $ {\bf LaAlO_3} $/$ {\bf SrTiO_3} $ heterostructures: \\
       electronic properties and impact on surface and interface 
       reconstruction} 

\author{I.~I.~Piyanzina} 
\affiliation{Center for Electronic Correlations and Magnetism, 
             Institute of Physics, University of Augsburg, 
             86135 Augsburg, Germany}
\affiliation{Institute of Physics, Kazan Federal University, 
             420008 Kazan, Russia}
\author{V.~Eyert}
\affiliation{Materials Design SARL, 
             42 Avenue Verdier, 92120 Montrouge, France}
\author{Yu.~V.~Lysogorskiy}
\affiliation{Institute of Physics, Kazan Federal University, 
             420008 Kazan, Russia}
\author{D.~A.~Tayurskii}
\affiliation{Institute of Physics, Kazan Federal University, 
             420008 Kazan, Russia}
\author{T.~Kopp}
\affiliation{Center for Electronic Correlations and Magnetism, 
             Institute of Physics, University of Augsburg, 
             86135 Augsburg, Germany}

\date{\today}

\begin{abstract}
We investigate the effect of oxygen vacancies and hydrogen dopants 
at the surface and inside slabs of $ {\rm LaAlO_3} $, $ {\rm SrTiO_3} $,  
and $ {\rm LaAlO_3} $/$ {\rm SrTiO_3} $ heterostructures on the 
electronic properties by means of electronic structure calculations as 
based on density functional theory. Depending on the concentration, 
the presence of these defects in a $ {\rm LaAlO_3} $ slab can suppress 
the surface conductivity. In contrast, in insulating $ {\rm SrTiO_3} $ 
slabs already very small concentrations of oxygen vacancies or hydrogen 
dopant atoms induce a finite occupation of the conduction band. Surface 
defects in insulating $ {\rm LaAlO_3} $/$ {\rm SrTiO_3} $ heterostructure 
slabs with three $ {\rm LaAlO_3} $ overlayers lead to the emergence of 
interface conductivity. Calculated defect formation energies reveal 
strong preference of hydrogen dopant atoms for surface sites for all 
structures and concentrations considered. Strong decrease of the defect 
formation energy of hydrogen adatoms with increasing thickness of the 
$ {\rm LaAlO_3} $ overlayer and crossover from positive to negative 
values, taken together with the metallic conductivity induced by hydrogen 
adatoms, seamlessly explains the semiconductor-metal transition observed  
for these heterostructures as a function of the overlayer thickness. 
Moreover, we show that the potential drop and concomitant shift of 
(layer resolved) band edges is suppressed for the metallic configuration.  
Finally, magnetism with stable local moments, which form atomically 
thin magnetic layers at the interface, is generated by oxygen vacancies 
either at the surface or the interface, or by hydrogen atoms buried 
at the interface. In particular, oxygen vacancies in the $ {\rm TiO_2} $ 
interface layer cause drastic downshift of the $ 3d $ $ e_g $ states of 
the Ti atoms neighboring the vacancies, giving rise to strongly localized 
magnetic moments, which add to the two-dimensional background magnetization. 
\end{abstract}


\maketitle

\section{Introduction}
\label{intro}

Oxide heterostructures have been in the focus of solid state research 
for more than a decade.\cite{mannhart2010,zubko2011,bibes2011} The 
intense exploration of the multilayered systems is well justified as 
electronic reconstruction\cite{hesper2000,okamoto2004,comment1} at 
the internal interfaces and the surfaces allows for a plethora of 
phenomena which are otherwise not observed in a single compound. 
Moreover, the sensitivity of competing or even coexisting phases to 
external fields and control parameters in the transition metal oxides 
makes the oxide heterostructures ideal candidates for multifunctional 
devices -- especially since metallic phases may be confined to thin
sheets with a thickness of a few lattice constants. 

For the paradigmatic oxide heterostructure with $ {\rm LaAlO_3} $ 
(LAO) thin films on $ {\rm SrTiO_3} $ (STO) 
substrates,\cite{ohtomo2004,nakagawa2006} distinct electronic phases 
have been extensively characterized at the LAO-STO interface: for LAO 
films with more than 3 layers\cite{thiel2006} and LaO termination 
towards the TiO$_2$ interface, a nm-wide\cite{sing2009} 
metallic\cite{ohtomo2004,thiel2006} state is formed in the STO layers 
next to the interface which becomes superconducting below a 
temperature of approximately 300~mK.\cite{reyren2007} Remarkably, the 
superconducting state coexists with a magnetic state possibly formed 
in patches of an inhomogeneous interface state.\cite{luli2011a,moler2011} 
The magnetism\cite{brinkman2007} appears 
to be stable up to room temperature but its origin has not been settled. 
\cite{pavlenko2012a,pavlenko2012b,michaeli2012,kalisky2012,pavlenko2013,lechermann2014,ruhman2014,yu2014} 
It may be well related to oxygen vacancies 
\cite{pavlenko2012b,park2013,salluzzo2013,pavlenko2013,lechermann2014}
at the interface and/or the surface which lead to an orbital 
reconstruction of nearby Ti sites and generate a local spin polarization. 
The orbital reconstruction involves a splitting of the Ti $ 3d $ $ e_g $ 
states on account of an oxygen vacancy in the vicinity of the Ti 
interface site. The $ e_g $ orbital with a lobe in the direction of 
the vacancy position is lowered energetically so substantially that 
it is partially occupied and Hund’s coupling induces magnetic moment 
formation. \cite{pavlenko2012b}
Other scenarios such as the formation of Ti$^{3+}$-on-Al$^{3+}$ defects 
in LAO near the interface\cite{yu2014} or the Zener exchange between 
an insulating interface layer and the nearest TiO$_2$ plane have been 
suggested.\cite{michaeli2012,banerjee2013} Importantly, Rashba spin-orbit 
coupling is generated at the interface\cite{caviglia2010,zhong2013} 
through inversion symmetry breaking which leads to anomalous 
magnetotransport\cite{caviglia2010,joshua2013,dagan2009,fuchs2015,seiler2016} 
and may even stabilize non-trivial topological 
states.\cite{fidkowski2013,loder2015,scheurer2015,kuerten2017} Moreover, 
capacitance measurements reveal that the electronic compressibility at the 
interface can be negative for low charge carrier concentration 
(controlled by a sufficiently large negative bias).\cite{luli2011b} It is 
not yet decided if the negative compressibility originates from interaction
effects,\cite{luli2011b,freericks2016,steffen2016} strong spin-orbit 
coupling,\cite{grilli2012,steffen2015} or from the electronic 
reconstruction in multiple planes.\cite{scopigno2016} 
Eventually, the dielectric properties of the heterostructure
control electronic  transport, and strain has a strong impact on 
them.\cite{zabaleta2016} Electrostrictive and flexo\-electric coupling 
on the STO side may play an important role in explaining the abruptness 
of the Lifshitz transition between low and intermediate electron density 
at the interface.\cite{raslan2018} Electronic devices 
from oxide interfaces are now feasible: circuits with all-oxide field-effect 
transistors were fabricated from LAO/STO heterostructures.\cite{forg2012} 
Recent progress in the fabrication of multilayered systems allows to 
manufacture stable polar nanostructures~\cite{noguera2008} and to make 
use of their respective functionalities.

With all these striking findings the electronic reconstruction at 
the LAO-STO interface is not yet profoundly understood. The polar
catastrophe mechanism, in which half an electron per planar unit 
cell is transferred from the surface to the interface, can neither 
account for the insulating surface state \cite{thiel2006,berner2013a} 
nor the sizable suppression of the internal electrostatic field in 
the LAO film \cite{segal2009,slooten2013,berner2013b} nor the built-up 
of a similar (superconducting) metallic interface state in samples 
where the polar LAO was replaced by amorphous aluminium oxide.
\cite{fuchs2014} 

In this respect, the question arises if defect states play an important 
role, especially if they are electron donor states.
\cite{bristowe2011,pavlenko2012b,janotti2012,berner2013b,bristowe2014,yu2014,krishnaswamy2015}
Also the debate concerning the origin of magnetism motivates to investigate 
defect states. Although defects are often considered to be unavoidable 
``dirt'' they nevertheless can decisively influence the electronic 
reconstruction at surfaces and interfaces and thereby contribute to 
their functionalization. However, we are still lacking a comprehensive 
account of the electronic reconstruction in the presence of defects. 
Certainly, this is beyond the scope of a single investigation. Here, we 
resort to first principles calculations and focus on oxygen vacancies 
and hydrogen dopants with the motivation that they are omnipresent and 
are both electron donors. The defect states have been analyzed before and 
a good understanding about their role in forming a conducting LAO-STO 
interface in conjunction with an insulating LAO surface is being 
developed.\cite{bristowe2011,pavlenko2012b,janotti2012,bristowe2014,yu2014,krishnaswamy2015} 
Here, we address defect profiles through the entire heterostructure 
and reconsider orbital reconstruction of the Ti $3d$ states with the 
formation of sizable magnetic moments. The essential issue is to understand 
the impact of defects on the interface electronic states. 

The paper is organized as follows: In Sec.~\ref{method} we briefly 
discuss our approach to model the perovskite slabs as well as the 
LAO-STO heterostructure and introduce the computational method 
that we employ in the presence of defects. We present the results on the 
electronic structure and defect formation energies in dependence on 
hydrogen adatoms and oxygen vacancies in LAO and STO slabs, and in 
LAO-STO heterostructures in subsections~\ref{lao}, \ref{sto} and,  
\ref{heterostructure}, respectively. In the last part of 
\ref{heterostructure}, we discuss the consequences of  surface passivation 
through hydrogen atoms, in particular the appearance of an insulating 
surface state in the presence of a conducting interface, the absence of 
a potential built-up in the LAO overlayers and a critical thickness of 
the LAO film. In Sec.~\ref{magnetism} we review the formation of local 
magnetic moments and orbital reconstruction through oxygen vacancies and
introduce a magnetic state with hydrogen defects at the the interface of 
the LAO-STO heterostructure. Sec.~\ref{summary} summarizes our results.

\section{Computational Method}
\label{method}

The {\it ab initio} calculations were based on density functional 
theory (DFT).\cite{hohenberg1964,kohn1965}
Exchange and correlation effects were accounted for by the generalized
gradient approximation (GGA) as parametrized by Perdew, Burke, and
Ernzerhof (PBE).\cite{perdew1996} The Kohn-Sham equations were solved
with projector-augmented-wave (PAW) potentials and wave functions
\cite{bloechl1994paw} as implemented in the Vienna Ab-Initio Simulation
Package (VASP),\cite{kresse1996a,kresse1996b,kresse1999} which is part of 
the MedeA\textsuperscript{\textregistered} software of Materials Design.
\cite{medea} Specifically, we used a plane-wave cutoff of 400~eV.
The force tolerance was 0.05~eV/\AA\ and the energy tolerance for the
self-consistency loop was $ 10^{-5} $~eV. The Brillouin zones were 
sampled using Monkhorst-Pack grids \cite{monkhorst1976} including 
$ 5 \times 5 \times 1 $ $ {\bf k} $-points in case of the $ 2 \times 2 $ 
supercells of $ {\rm LaAlO_3} $ and $ {\rm SrTiO_3} $ slabs. In contrast, 
for the heterostructures with their increased number of atoms we used 
$ 3 \times 6 \times 1 $ $ {\bf k} $-points for the $ 2 \times 1 $ 
supercells and $ 3 \times 3 \times 1 $ $ {\bf k} $-points for the 
$ 2 \times 2 $ supercells. Finally, the electronic densities of states 
were calculated using the linear tetrahedron method \cite{bloechl1994ltm} 
on $ 4 \times 4 \times 1 $ $ {\bf k} $-point grids. 

All calculations were performed in the framework of the GGA+$ U $ 
method using the simplified approach proposed by Dudarev {\it et al.}, 
\cite{dudarev1998} which takes only the difference of $ U - J $ into 
account. We applied additional local correlations of 
$ U - J = 8 $\,eV and $ U - J = 2 $\,eV to the La $ 4f $ and the 
Ti $ 3d $ orbitals, respectively. This choice is based on our previous 
detailed investigation of the effect of local electronic correlations 
as captured by the GGA+$ U $ method.\cite{piyanzina2017} In that study 
we performed a systematic variation of the $ U - J $ values for the 
aforementioned two orbitals and their effect on the atomic and electronic 
structure, which led us to conclude that the above values represent good 
choices and thereby allow for an accurate and at the same time efficient 
assessment of the systems under study. This conclusion was also supported 
by our previous work using the GGA+$ U $ method and the hybrid functional 
approach.\cite{breitschaft2010,cossu2013} 

In order to study the $ {\rm LaAlO_3} $~and $ {\rm SrTiO_3} $~surfaces,  
slabs were constructed comprising $ 6 \frac{1}{2} $ unit cells stacked 
along the [001] direction and identical termination on both sides 
($ {\rm AlO_2} $ for $ {\rm LaAlO_3} $ and $ {\rm TiO_2} $ for 
$ {\rm SrTiO_3} $). While the atomic positions of the outer parts of 
these slabs were allowed to relax, the central region of $ 1 \frac{1}{2} $ 
unit cells was kept fixed in order to simulate the bulk structure. 
\cite{krishnaswamy2014,lin2009}

The heterostructures were modeled by a central region of $ {\rm SrTiO_3} $
comprising $ 4 \frac{1}{2} $ unit cells with $ {\rm TiO_2} $ termination 
on both sides and three $ {\rm LaAlO_3} $ overlayers with $ {\rm LaO} $
termination towards the central slab and $ {\rm AlO_2} $ surface
termination also on both sides (see Fig.~4 of Ref.~\onlinecite{piyanzina2016}).
This slab model guaranteed a non-polar structure without any artificial
dipoles. Finally, in order to avoid interaction of the surfaces and slabs
with their periodic images, 20~\AA-wide vacuum regions were added in accordance
with previous work. \cite{cossu2013,piyanzina2016} The in-plane lattice
parameter $ a = b = 3.905 $\,\AA\ was fixed to the experimental values of
bulk $ {\rm SrTiO_3} $ \cite{ohtomo2004} and kept for all subsequent
calculations reflecting the stability of the substrate. However, in 
contrast to the pure LAO and STO slabs mentioned in the previous 
paragraph in the heterostructure with its thinner STO core the atomic 
positions of all atoms were fully relaxed. 

In passing we mention that throughout in this paper the term heterostructure 
refers to slab models of the kind just described, in particular, with a wide 
vacuum region separating the slabs, rather than a structures with infinite 
alternation of $ {\rm LaAlO_3} $ and $ {\rm SrTiO_3} $ slabs. 

Finally, to study the influence of defects we used $ 1 \times 1 $, 
$ 1 \times 2 $ and $ 2 \times 2 $ in-plane supercells of the above 
cells. Both oxygen vacancies and hydrogen dopants were introduced 
at the surface or in a subsurface layer on both sides while preserving 
the inversion symmetry of the slabs. Defect formation energies were 
calculated following the standard procedure adopted by many authors. 
\cite{zhang1991,reuter2001,li2011,zhang2010,son2010,freysoldt2014}
Specifically, the oxygen vacancy formation energy $ E_{\rm form} $ is 
defined as
\begin{equation}
E_{\rm form} = \left( E_{\rm total} 
                      - \left( E_{\rm bare} - n \mu_{\rm O} (T, p) \right)
               \right)/n \,,
\label{eq:eform1}
\end{equation}
where $ n $ is the number of vacancies per unit cell. $ E_{\rm bare} $ 
and $ E_{\rm total} $ are the total energies of the bare and 
oxygen-deficient slab, respectively. Finally, $ \mu_{\rm O} (T, p) $ 
denotes the chemical potential of oxygen, which depends on temperature 
and oxygen partial pressure. However, the oxygen chemical potential 
is confined by lower and upper limits reflecting the stability of the 
system involved. In particular, in the ``oxygen-poor limit'', where the 
chemical potential is low, oxygen would start to dissociate and thereby 
destabilize the slab. In contrast, in the ``oxygen-rich limit'', oxygen 
would start to form a condensed $ {\rm O_2} $ solid phase covering the 
surface. At ambient conditions both limits are out of reach and a 
resonable choice for the upper limit is 
\begin{equation}
\mu_{\rm O} (T, p) = \frac{1}{2} E_{\rm O_2} 
                        \,,
\label{eq:eform2}
\end{equation}
where $ E_{\rm O_2} $ is the total energy of a free, isolated 
$ {\rm O_2} $ molecule at $ T = 0 $\,K.
\cite{zhang1991,reuter2001,li2011,zhang2010,son2010,freysoldt2014,silva2016}.  
Note that the total energy of the oxygen molecule was determined 
from spin-polarized calculations. 

Similarly, the hydrogen-dopant formation energy $ E_{\rm form} $ 
is given by 
\begin{equation}
E_{\rm form} = \left( E_{\rm total} 
                      - \left( E_{\rm bare} + n \mu_{\rm H} (T, p) \right)
               \right)/n \,,
\label{eq:eform3}
\end{equation}
where $ n $ is the number of dopant atoms per unit cell. $ E_{\rm total} $  
is the total energy of the slab with dopant atoms symmetrically added 
on both sides, whereas $ \mu_{\rm H} (T, p) $ denotes the chemical potential 
of hydrogen. As for the case of oxygen, the latter is restricted to a 
certain range in order to preserve the thermodynamic stability of the slab. 
Again following the literature, we here resort to the choice 
\begin{equation}
\mu_{\rm H} (T, p) = \frac{1}{2} E_{\rm H_2} 
                        \,,
\label{eq:eform4}
\end{equation}
where $ E_{\rm H_2} $ is the total energy of a free, isolated 
$ {\rm H_2} $ molecule at $ T = 0 $\,K obtained from spinpolarized 
calculations.  
\cite{zhang1991,reuter2001,li2011,zhang2010,son2010,freysoldt2014,silva2016}.  

In the present context it is important to note that although the 
aforementioned defects may not even be thermodyamically stable at 
experimentally accessible conditions, they may nevertheless be 
created during the synthesis of samples and kinetically frozen. 
Yet, the above relations may still serve as a means to access the 
relative stability of defects located at different layers.

\section{Impact of impurities on structures and electronic properties} 
\label{impurities}

\subsection{$ {\bf LaAlO_3} $ slab}
\label{lao}

Bulk $ {\rm LaAlO_3} $ is a wide gap insulator with a band gap of 
5.6\,eV and crystallizing in the perovskite structure. \cite{ohtomo2004} 
It is used as a high-$\kappa$ gate dielectric for the recessed-gate 
GaN MOSFET transistors, \cite{tsai2012} as a substrate 
for high-$T_{\textit{c}}$ superconductors, \cite{mcintyre1994} and it 
exhibits colossal magnetoresistance. \cite{wang1996} Recent findings 
suggest a two-dimensional electron gas at the $ {\rm LaAlO_3} $ surface, 
depending on the localization of electrons or holes generated by defects. 
\cite{silva2016}

\begin{figure}[tb] %%LAO structure
\centering
\includegraphics[width=1.0\linewidth]{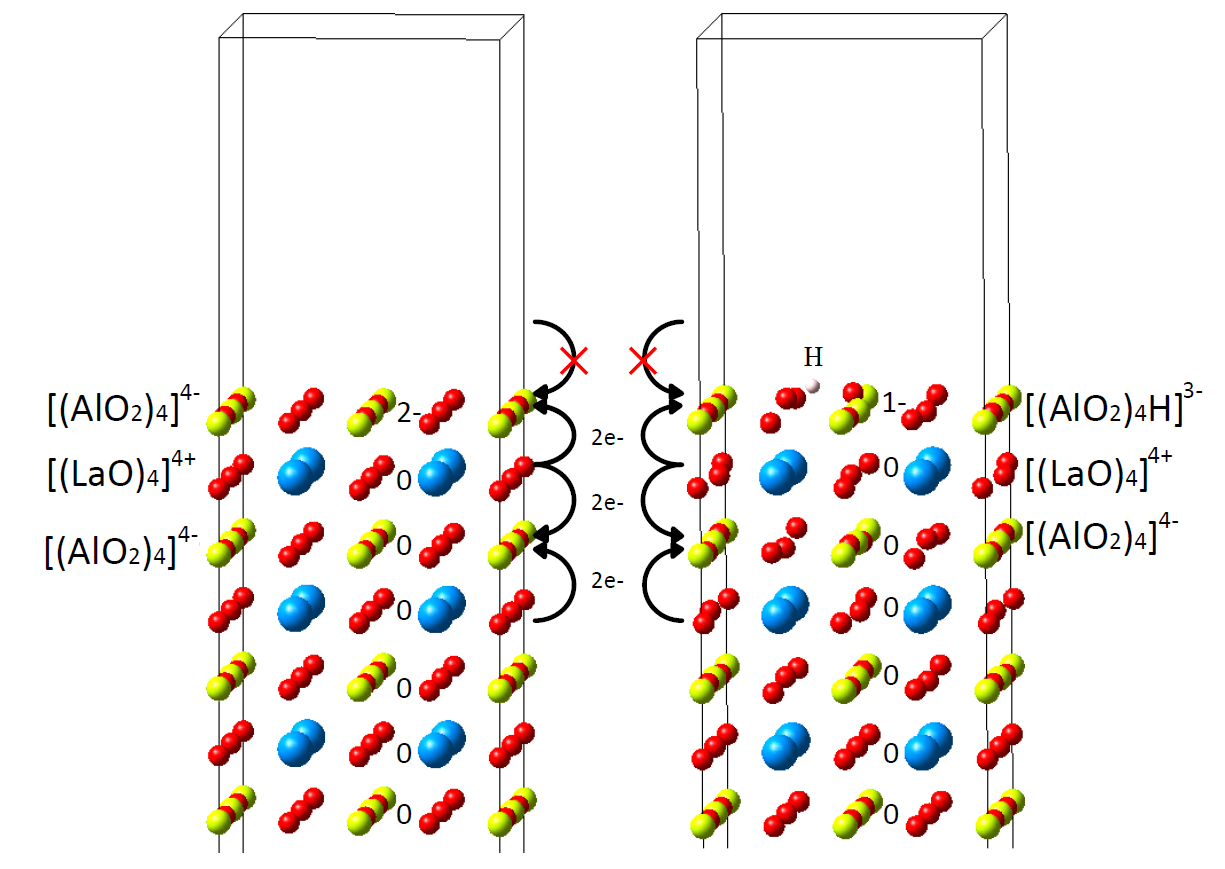}
\caption{Structures and ionic models of a bare $ {\rm LaAlO_3} $ slab 
         (left) and a slab with one hydrogen adatom per $ 2 \times 2 $ 
         surface unit cell (right). In both cases the surface is charged 
         but the charge concentrations differ. In contrast, one oxygen 
         vacancy or two hydrogen atoms (per $ 2 \times 2 $ surface cell) 
         will result in full charge compensation and insulating states. 
         Only the upper part of the slab is displayed. Arrows indicate 
         electron transfer from LaO planes to the neighboring 
         $ {\rm AlO_2} $ planes. The surfaces lack two (bare surface) 
         and one (for one H adatom) electron. The formal charge transfers 
         of the $ 2 \times 2 $ stacked planar cells are marked in the 
         center of the plot between the two structures.} 
\label{fig:lao-charge}
\end{figure}

\begin{figure}[tb] %%LAO DOSes
\centering 
\subfigure{(a)
           \includegraphics[width=0.25\linewidth]{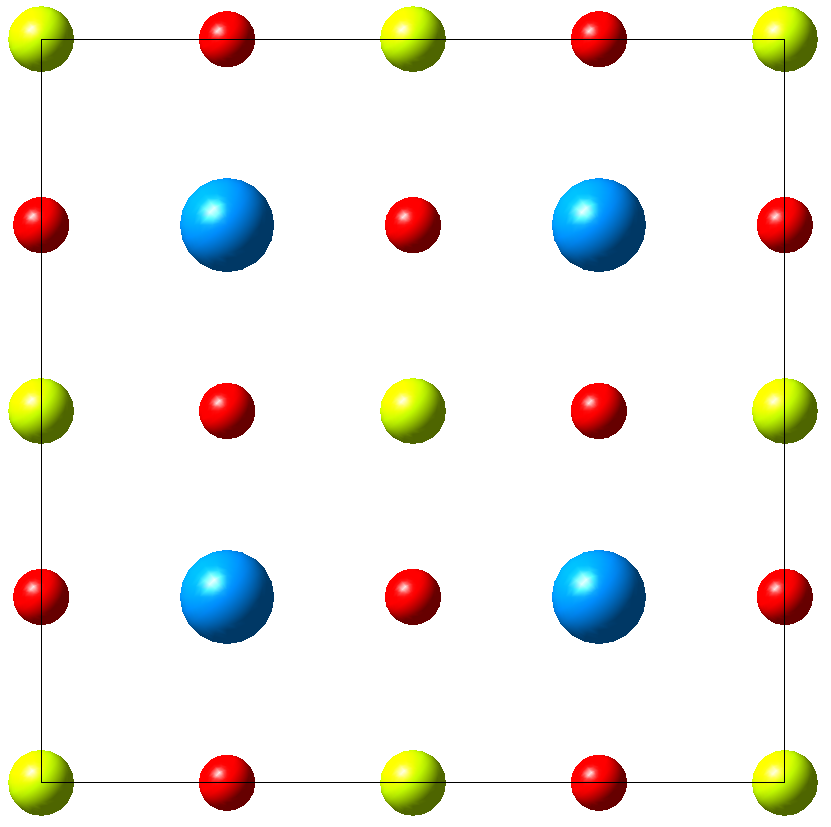}
           \includegraphics[width=0.68\linewidth]{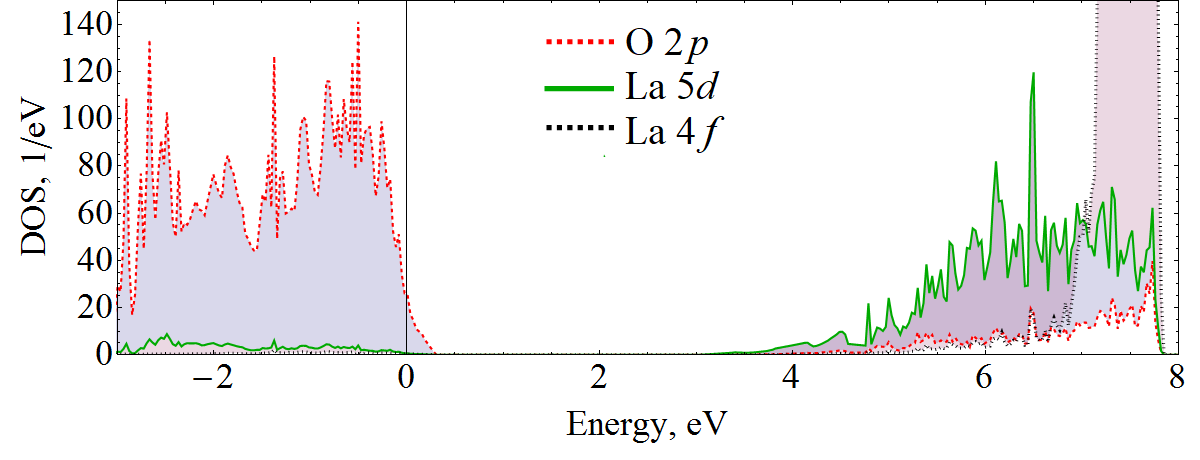}
           \label{fig:lao-dos-bare} }
\subfigure{(b)
           \includegraphics[width=0.25\linewidth]{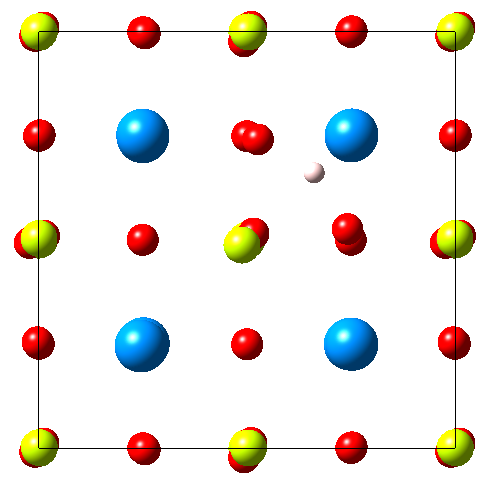}
           \includegraphics[width=0.68\linewidth]{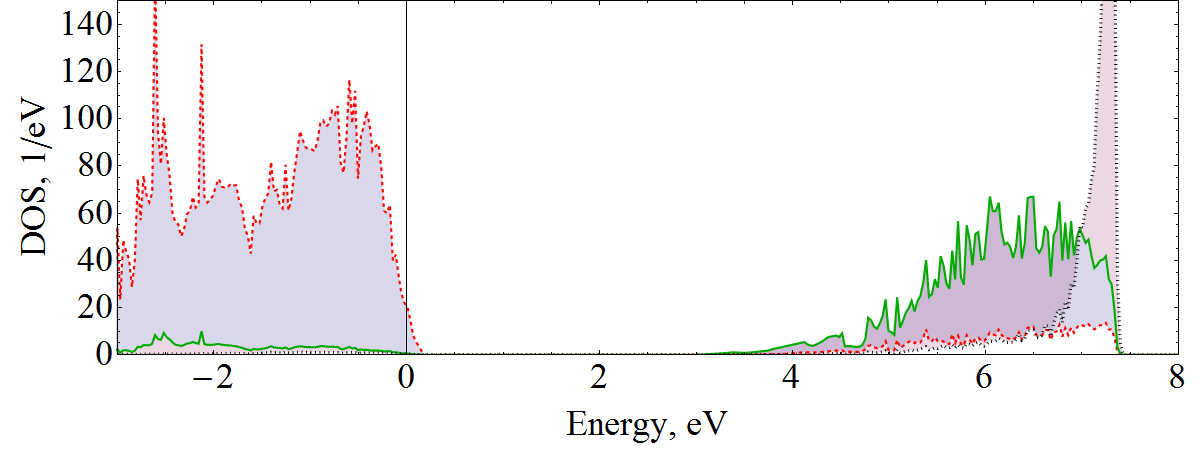}
           \label{fig:lao-dos-1H} }
\subfigure{(c)
           \includegraphics[width=0.25\linewidth]{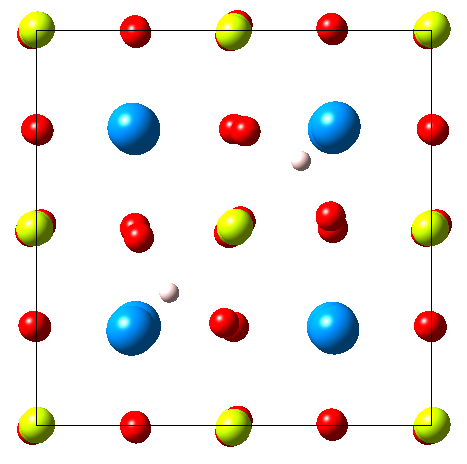}
           \includegraphics[width=0.68\linewidth]{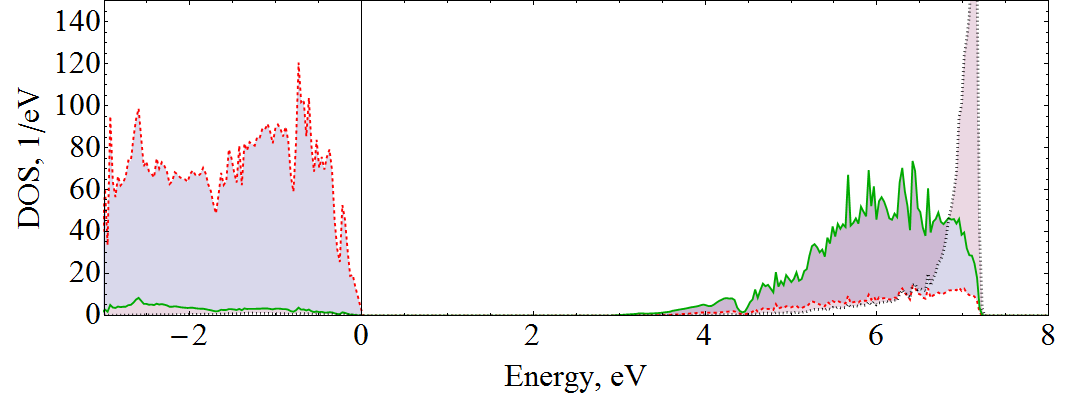}
           \label{fig:lao-dos-2H}} 
\subfigure{(d)
           \includegraphics[width=0.25\linewidth]{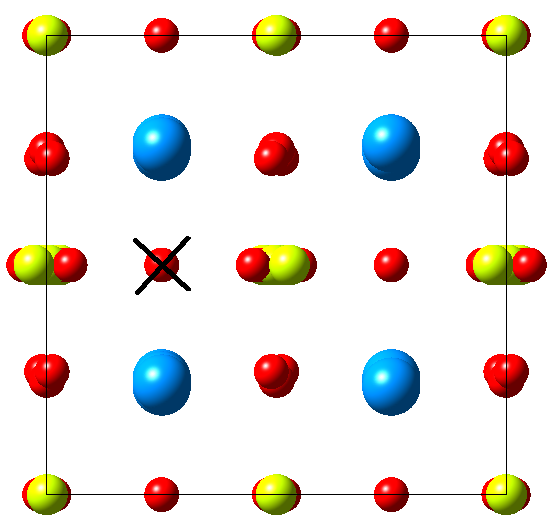}
           \includegraphics[width=0.68\linewidth]{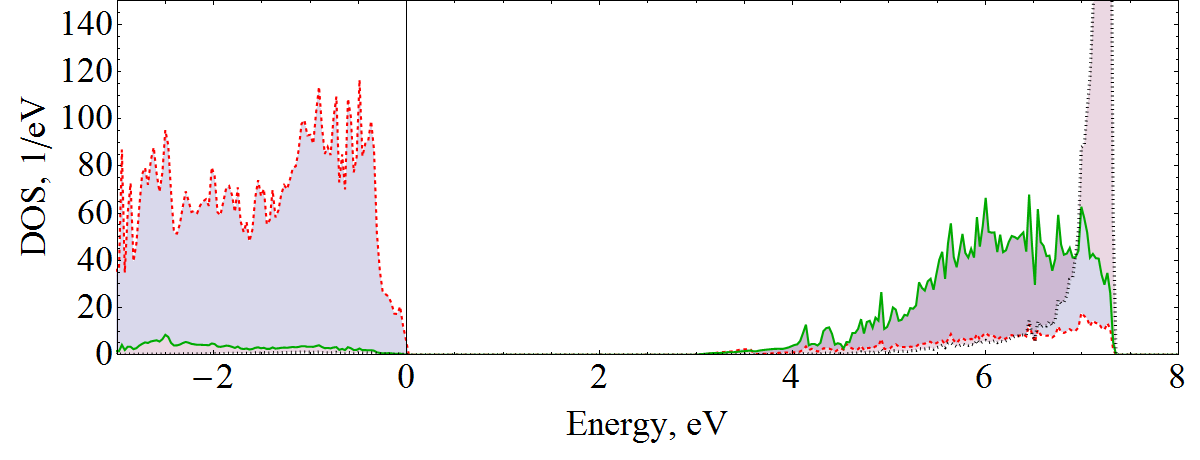}
           \label{fig:lao-dos-Ovac} }
\subfigure{(e)
           \includegraphics[width=0.25\linewidth]{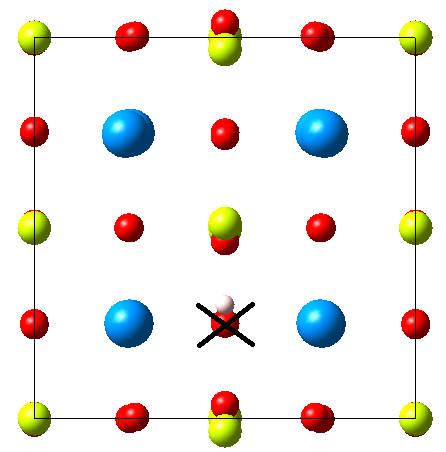}
           \includegraphics[width=0.68\linewidth]{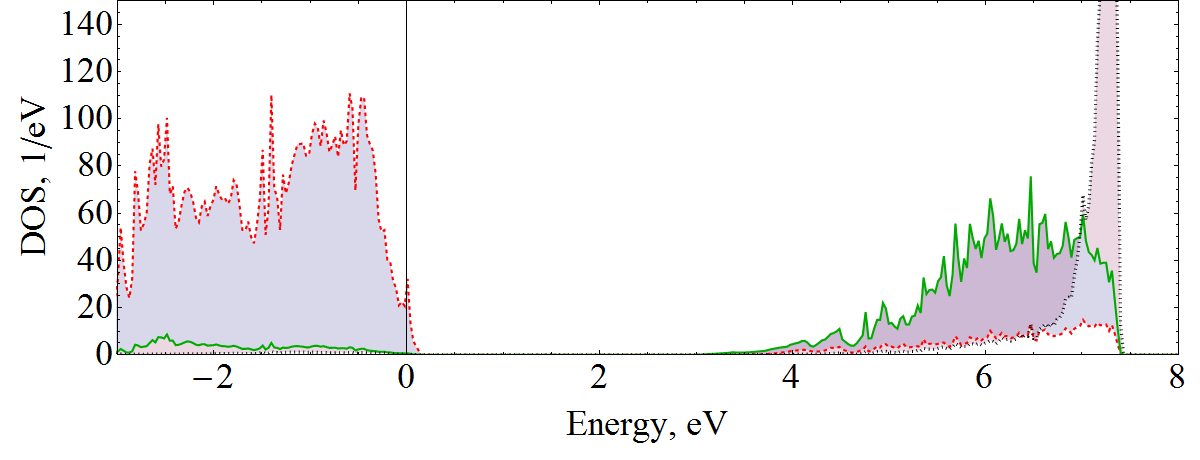}
           \label{fig:lao-dos-OandH} }
\caption{Left column: top views of \subref{fig:lao-dos-bare} bare 
                      $ 2 \times 2 $ $ {\rm LaAlO_3} $~surface as  
                      well as a surface with 
                      \subref{fig:lao-dos-1H} one hydrogen atom, 
                      \subref{fig:lao-dos-2H} two hydrogen atoms, 
                      \subref{fig:lao-dos-Ovac} an oxygen vacancy, 
                      and \subref{fig:lao-dos-OandH} an oxygen 
                      vacancy filled with a hydrogen atom. 
                      Aluminum, lanthanum, oxygen, and hydrogen atoms 
                      are given in yellow, blue, red, and light grey, 
                      respectively, whereas oxygen vacancies are marked 
                      by a cross. 
         Right column: Corresponding partial electronic densities of 
                       states.}
\label{fig:lao-dos}
\end{figure}

The structures of the bare $ {\rm AlO_2} $-terminated $ {\rm LaAlO_3} $ 
slab  and the same slab with a hydrogen adatom are shown in 
Fig.~\ref{fig:lao-charge}. 
The corresponding top views of $ 2 \times 2 $ surface supercells of the 
bare slab and slabs with different kinds of defects are displayed in 
Fig.~\ref{fig:lao-dos} 
together with the calculated partial electronic densities of states. 
Obviously, the bare $ {\rm LaAlO_3} $ surface, while still showing 
the large separation between O $ 2p $ and La $ 5d $ states, which 
causes the insulating behavior in the bulk material, exhibits 
metallic behavior due a small fraction of unoccupied O $ 2p $ bands 
as has been also found by Krishnaswamy {\it et al.}~\cite{krishnaswamy2014}
as well as by Silva and Dalpian.\cite{silva2016}  These states arise 
from the dangling bonds of the orbitals at the surface, which 
introduce a metallic surface layer. The situation can be understood 
within a simple ionic picture, where $ {\rm LaAlO_3} $ is considered 
as an alternating sequence of charged $ {\rm (AlO_2)^{-}} $ and 
$ {\rm (LaO)^{+}} $ layers. Thus, formally each $ {\rm LaO} $ layer 
provides half an electron to the $ {\rm AlO_2} $ layer below and 
above such that all atomic shells are completely filled. This ionic 
model is sketched in Fig.~\ref{fig:lao-charge} for a $ 2 \times 2 $ 
surface unit cell. Since at the surface this perfect compensation 
scheme is interrupted, the terminating $ {\rm AlO_2} $ layer misses 
two electrons (per $ 2 \times 2 $ surface unit cell) to completely 
fill its O $ 2p $ states and metallic conductivity equivalent to hole 
doping arises. 

The electronic charge in the surface layer needed to restore 
insulating behavior can be provided by two hydrogen adatoms or 
a single oxygen vacancy per $ 2 \times 2 $ surface unit cell. 
These two cases are displayed in Figs.~\ref{fig:lao-dos-2H} and 
\ref{fig:lao-dos-Ovac}. According to the corresponding calculated 
densities of states the proper amount of these defects thus leads 
to a complete filling of the O $ 2p $ states and recovery of the 
insulating behavior seen in bulk $ {\rm LaAlO_3} $ as has been 
also observed by Krishnaswamy and coworkers.\cite{krishnaswamy2014}  
The ionic model displayed in Fig.~\ref{fig:lao-charge} indicates 
that in the case of hydrogen adatoms the surface layer now can be 
formally written as $ {\rm [(AlO_2)_4H_2]^{2-}} $, whereas the 
layers below arise as alternating $ {\rm [(LaO)_4]^{4+}} $ and 
$ {\rm [(AlO_2)_4]^{4-}} $ as before. The hydrogen adatoms thus 
offer two electrons, which in a perfect crystal would be provided 
by the neighboring $ {\rm LaO} $ layer and thereby restore the 
insulating behavior at the surface. The same 
effect is achieved by a single oxygen vacancy. 

Finally, the case of only one hydrogen adatom per $ 2 \times 2 $ 
surface unit cell at the surface as shown in Fig.~\ref{fig:lao-dos-1H} 
or an oxygen vacancy filled by one hydrogen atom as displayed in 
Fig.~\ref{fig:lao-dos-OandH} are intermediate between the situations 
discussed above. Even though they both cause lowering of the O $ 2p $ 
dominated bands, these states are still not completely filled and metallic 
behavior remains. Within the ionic model the top layers can be formally 
written as $ {\rm [(AlO_2)_4H_2]^{1-}} $ in case of a single hydrogen 
adatom, i.e.\ these layers are missing one electron to completely fill 
the atomic O $ 2p $ shell. 

According to layer-resolved analysis of the partial densities of states 
of the bare $ {\rm LaAlO_3} $ surface the O $ 2p $ contribution at the 
Fermi energy drops from 10 states/eV in the surface layer to 2 states/eV 
in the second and third layer. In contrast, oxygen vacancies cause insulating 
behavior. However, with the vacancies filled by a single hydrogen atom 
the O $ 2p $ contribution at the Fermi energy drops from about 9 states/eV 
for the surface layer to 4 states/eV in the second and third layers 
indicating an increased thickness of the metallic surface layer as 
compared to the bare surface.

\begin{figure}[tb]  %%LAO defects
\vspace{4ex} 
\centering 
\subfigure{(a)
           \includegraphics[width=0.95\linewidth]{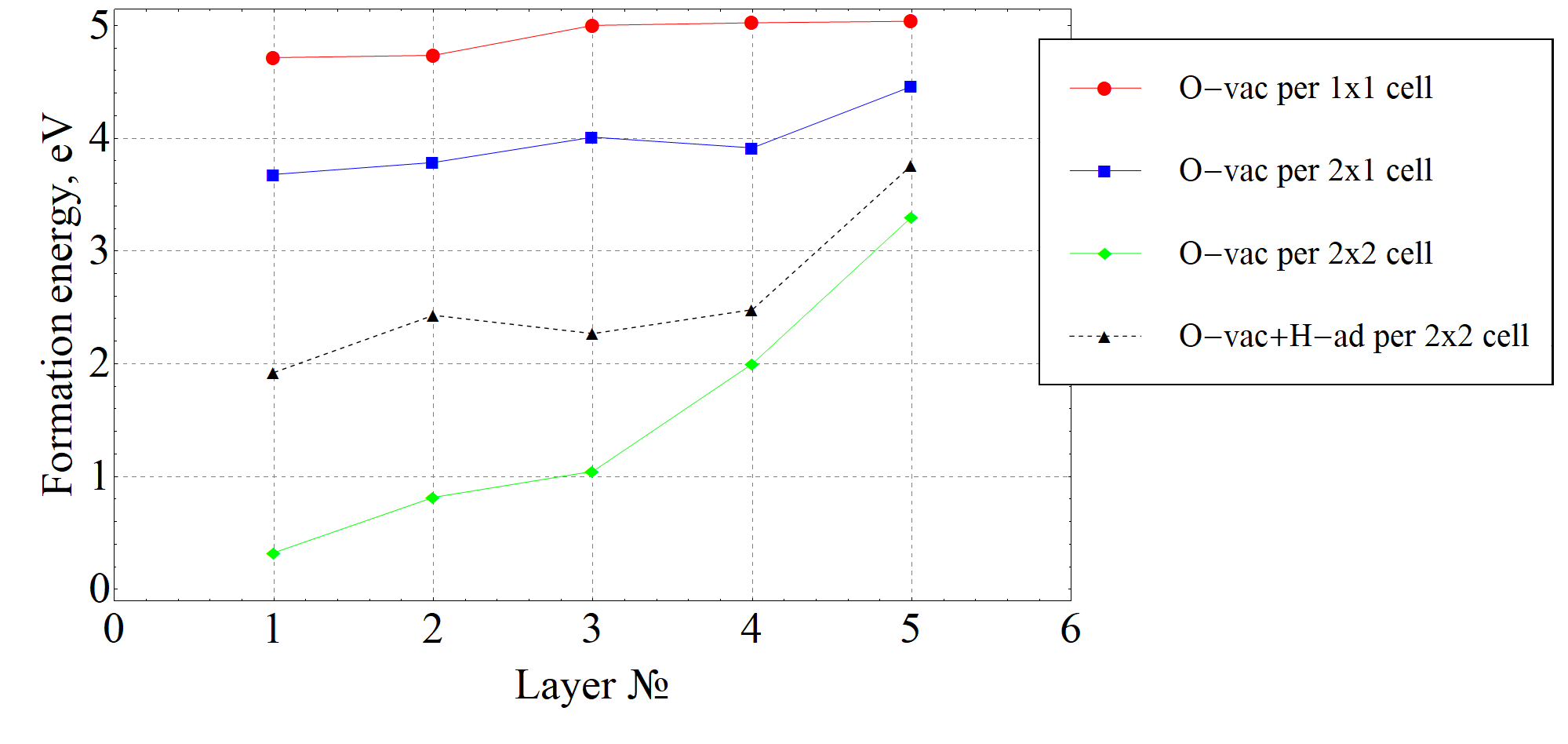} 
           \label{fig:lao-energy-O-vac-dep} }  
\subfigure{(b)
           \includegraphics[width=0.95\linewidth]{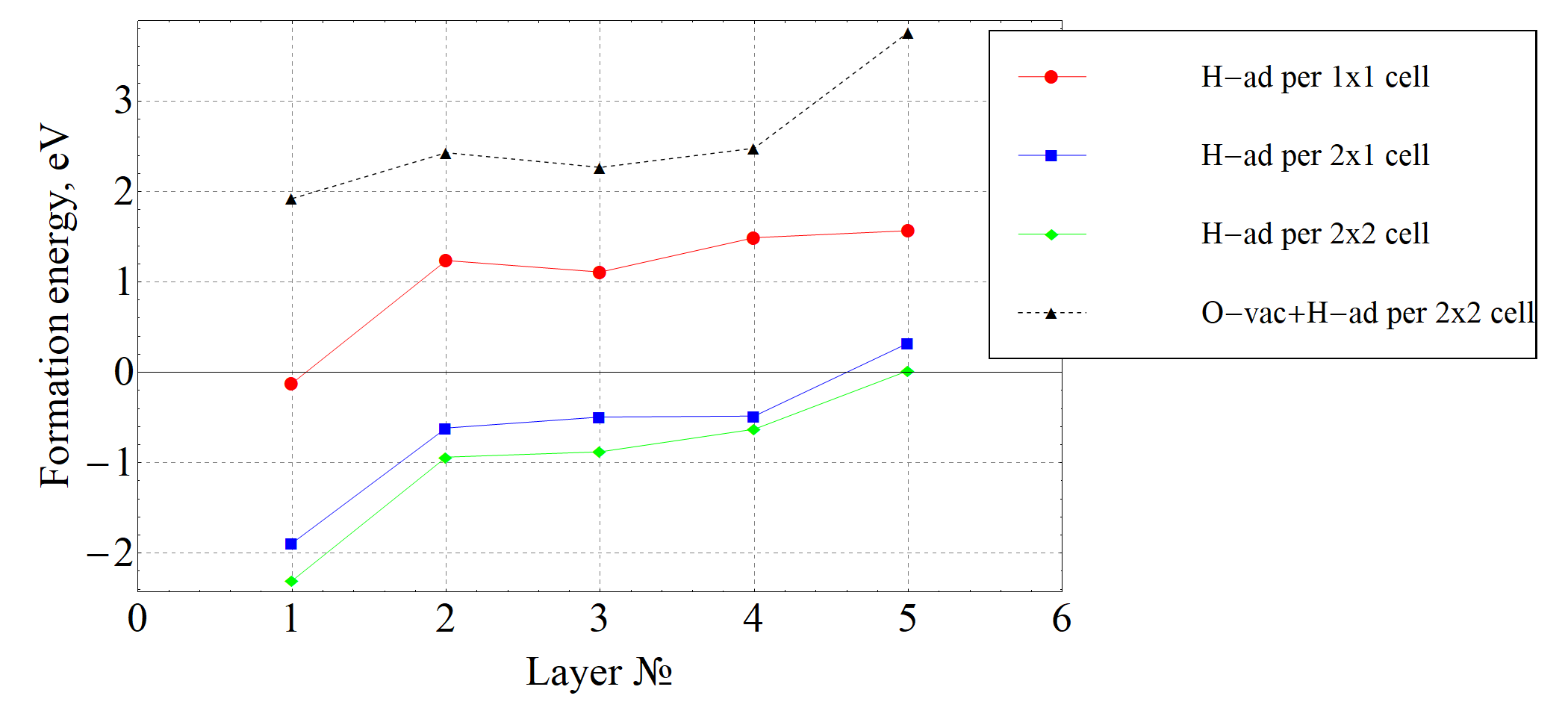} 
           \label{fig:lao-energy-H-ad-dep} }
\caption{Defect formation energies of $ {\rm AlO_2} $-terminated 
         $ {\rm LaAlO_3} $~slabs \subref{fig:lao-energy-O-vac-dep} 
         with an oxygen vacancy and \subref{fig:lao-energy-H-ad-dep} 
         with a hydrogen dopant atom in dependence of the position 
         of the defect. Layer 1 is the surface layer and counting 
         of layers treats LaO and $ {\rm AlO_2} $ layers separately.} 
\label{fig:lao-energy}
\end{figure}

Calculated energies of defect formation are displayed in 
Fig.~\ref{fig:lao-energy} 
for both oxygen vacancies and hydrogen dopant atoms at different 
concentrations as mimicked by different surface unit cells. In 
addition, the case of an oxygen vacancy filled with a single hydrogen 
atom was considered. The corresponding curve was included in both 
subfigures and serves as an orientation. All defects were placed at 
different layers ranging from the surface layer to the layer next to 
the center of the slab. Obviously, positions at the surface are preferred 
over positions well inside the slab. This trend is most clear for the 
case of an oxygen vacancy in a $ 2 \times 2 $ surface unit cell, which 
has been also considered by Silva and Dalpian.\cite{silva2016} Specifically, 
these authors find an increase of the formation energy of an oxygen vacancy 
from about 0.3 eV for a vacancy at the surface to about 3.5 eV for a 
vacancy in the fifth layer in almost perfect agreement with our results. 
Note that due to the Coulomb repulsion between neighboring defects 
the energies of defect formation show a strong increase with increasing 
defect concentration. Finally, we point to the negative formation energy 
for hydrogen dopant atoms at low concentrations indicating the likelihood 
of hydrogen adsorption restoring the insulating behavior at the surface.

\subsection{$ {\bf SrTiO_3} $ slab}
\label{sto}

\begin{figure}[t] %%STO structures with charges
\centering
\center{\includegraphics[width=1.0\linewidth]{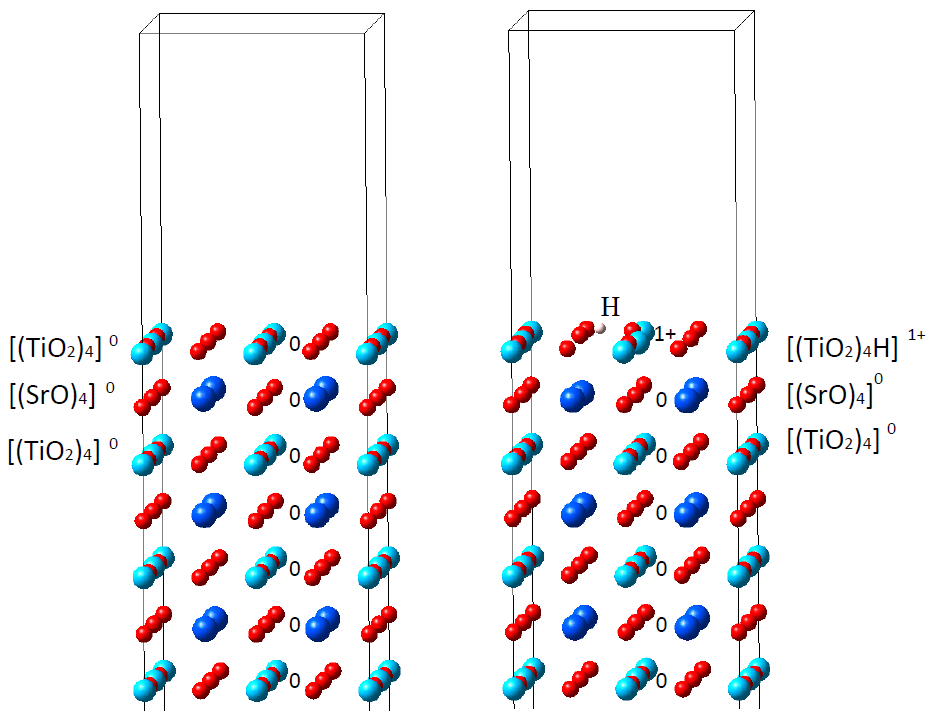}  \\}
\caption{Structures and ionic models of a bare $ {\rm SrTiO_3} $ slab 
         (left) and a slab with one hydrogen adatom per $ 2 \times 2 $ 
         surface unit cell (right). Within the ionic model there is no 
         charge transfer between layers and hydrogen adatoms and oxygen 
         vacancies at the surface act as electron donors causing finite 
         surface conductivity. Only the upper part of the slab is 
         displayed.} 
\label{fig:sto-charge}
\end{figure}

$ {\rm SrTiO_3} $ is an insulator with a gap of 3.2\,eV. Like 
$ {\rm LaAlO_3} $ it crystallizes in the perovskite structure and 
is widely used as a substrate for growing thin films as well as in 
electronic devices such as ceramic capacitors and variable resistors. 
However, these devices turned out to be affected by hydrogen adsorption 
due to the dissociation of adsorbed water.\cite{chen1998,lin2009} This 
tendency was confirmed by DFT calculations and surface metallicity 
induced by electron donation from hydrogen was suggested.\cite{lin2009} 
Oxygen vacancies can also affect the electrical properties. The 
emergence of a two-dimensional electron system at a vacuum-cleaved 
$ {\rm SrTiO_3} $ surface was observed in angle-resolved 
photoelectron spectroscopy.\cite{santander2011,meevasana2011} First 
principles calculations gave evidence that such a two-dimensional 
electron gas can be generated by oxygen vacancies at the 
surface.\cite{shen2012,silva2014}  
Moreover, electronic correlations were investigated within DFT+DMFT for 
oxygen-deficient $ {\rm SrTiO_3} $ surfaces.\cite{lechermann2016} 
Recently, the density of oxygen vacancies at $ {\rm SrTiO_3} $ surfaces 
was adjusted in situ during photoemission experiments.~\cite{dudy2016} 
Most prominently, electronic phase separation was identified with a highly 
doped metallic and insulating or lowly doped phase.~\cite{dudy2016} The 
authors concluded that, most likely, this imhomogeneity is driven chemically 
by clustering of oxygen vacancies.

\begin{figure}[tb] %%STO DOSes
\centering
\subfigure{(a)
           \includegraphics[width=0.25\linewidth]{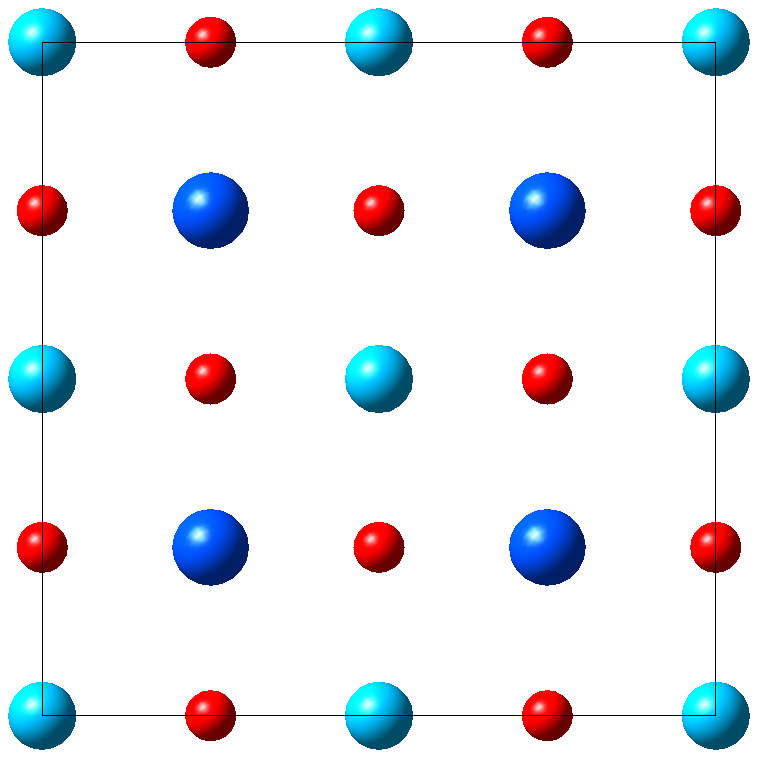}
           \includegraphics[width=0.68\linewidth]{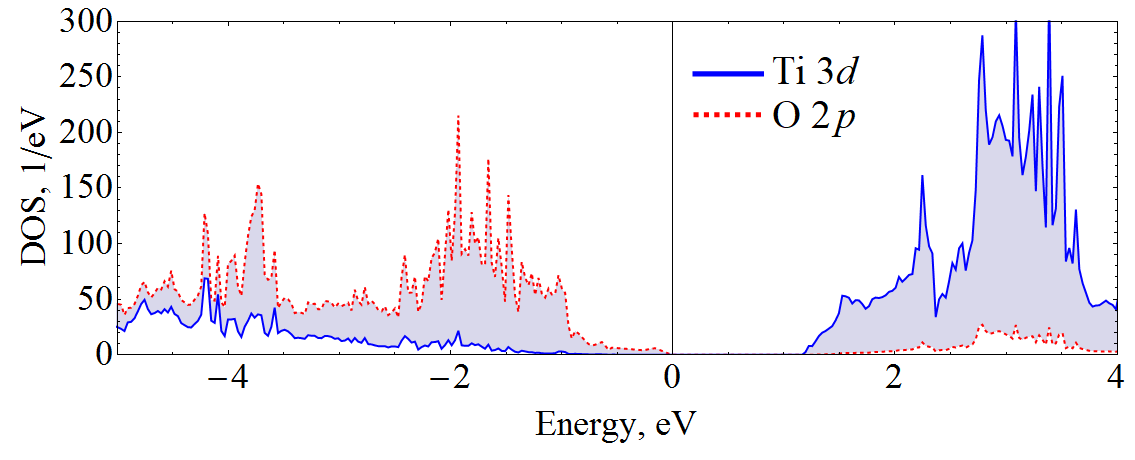}
           \label{fig:sto-dos-bare} }
\subfigure{(b)
           \includegraphics[width=0.25\linewidth]{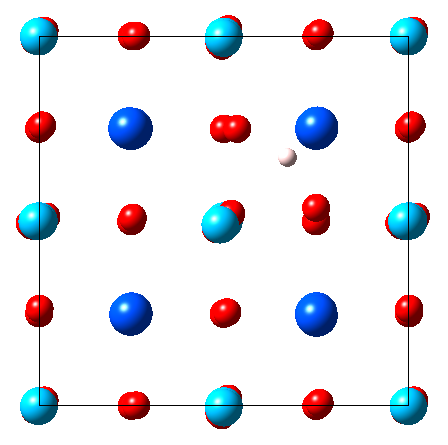}
           \includegraphics[width=0.68\linewidth]{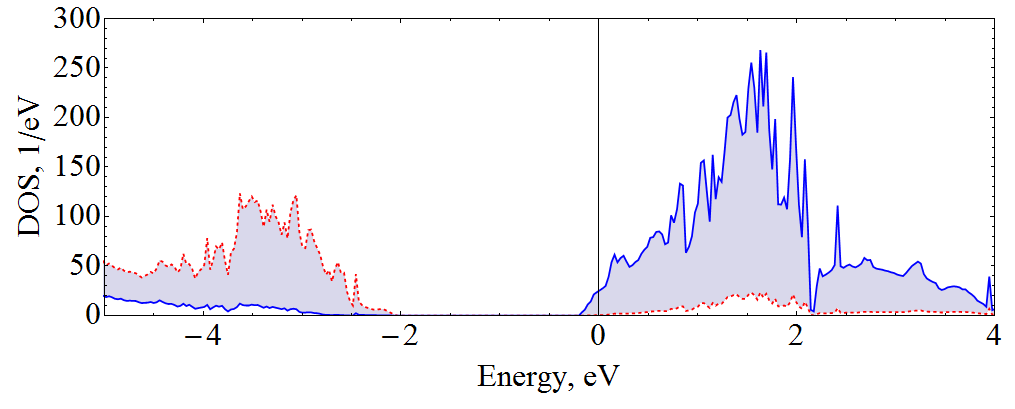}
           \label{fig:sto-dos-1H} }
\subfigure{(c)
           \includegraphics[width=0.25\linewidth]{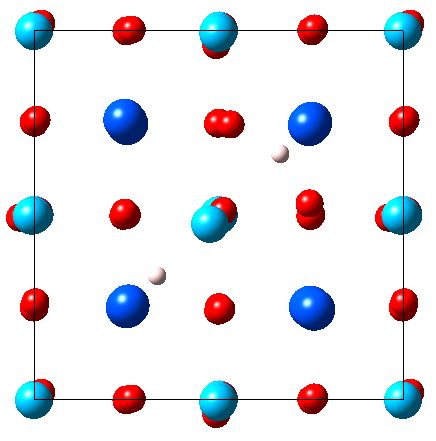}
           \includegraphics[width=0.68\linewidth]{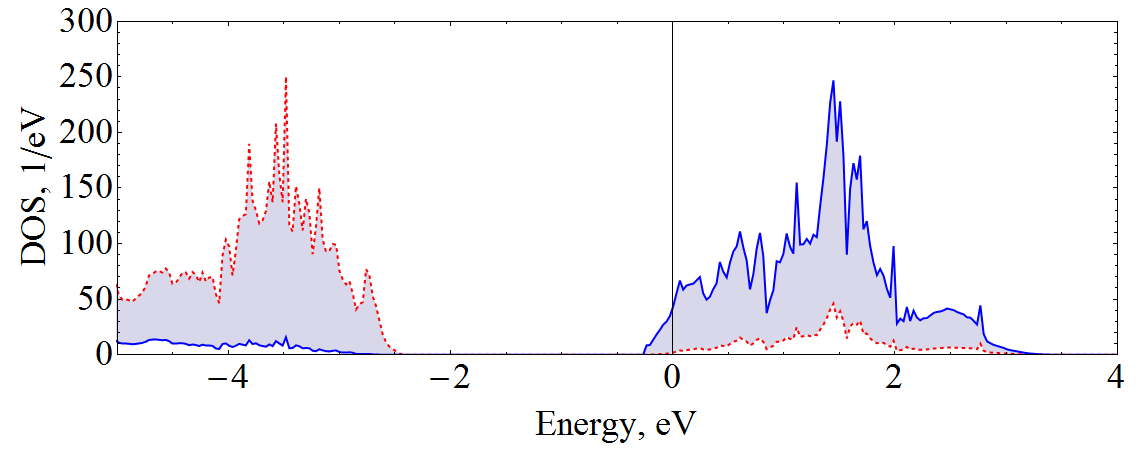}
           \label{fig:sto-dos-2H}} 
\subfigure{(d)
           \includegraphics[width=0.25\linewidth]{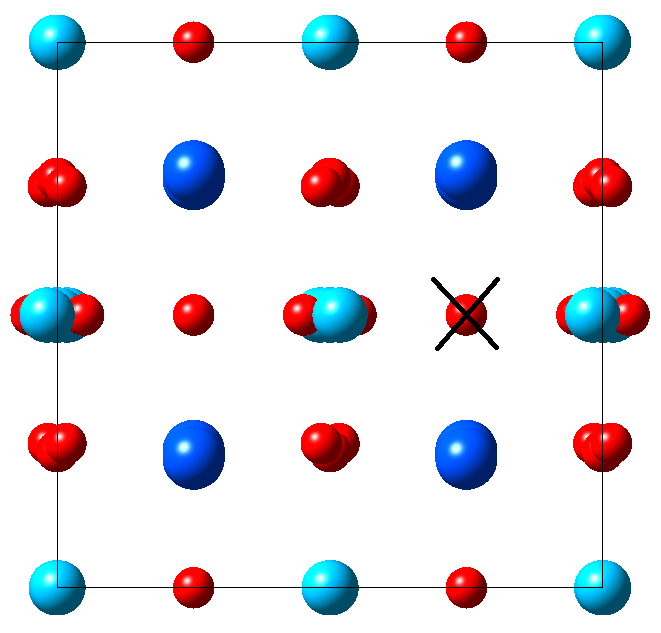}
           \includegraphics[width=0.68\linewidth]{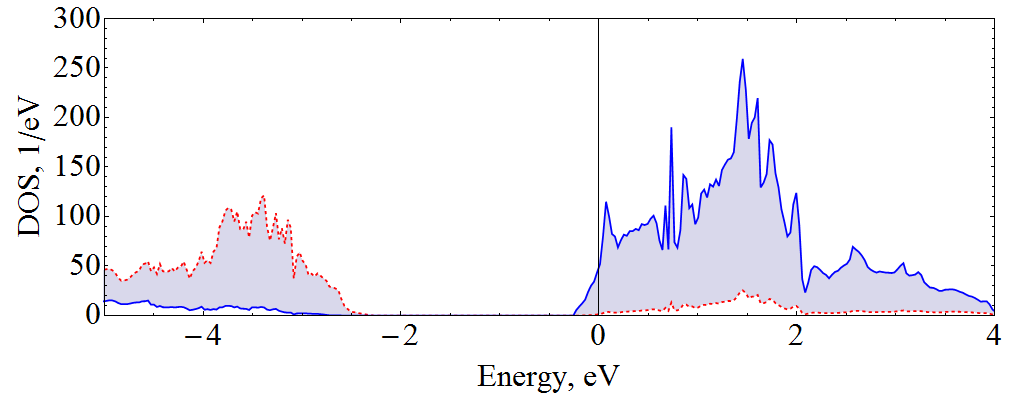}
           \label{fig:sto-dos-Ovac} }
\subfigure{(e)
           \includegraphics[width=0.25\linewidth]{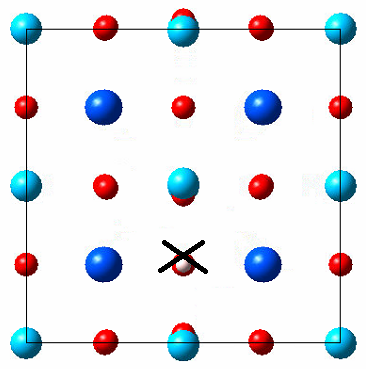}
           \includegraphics[width=0.68\linewidth]{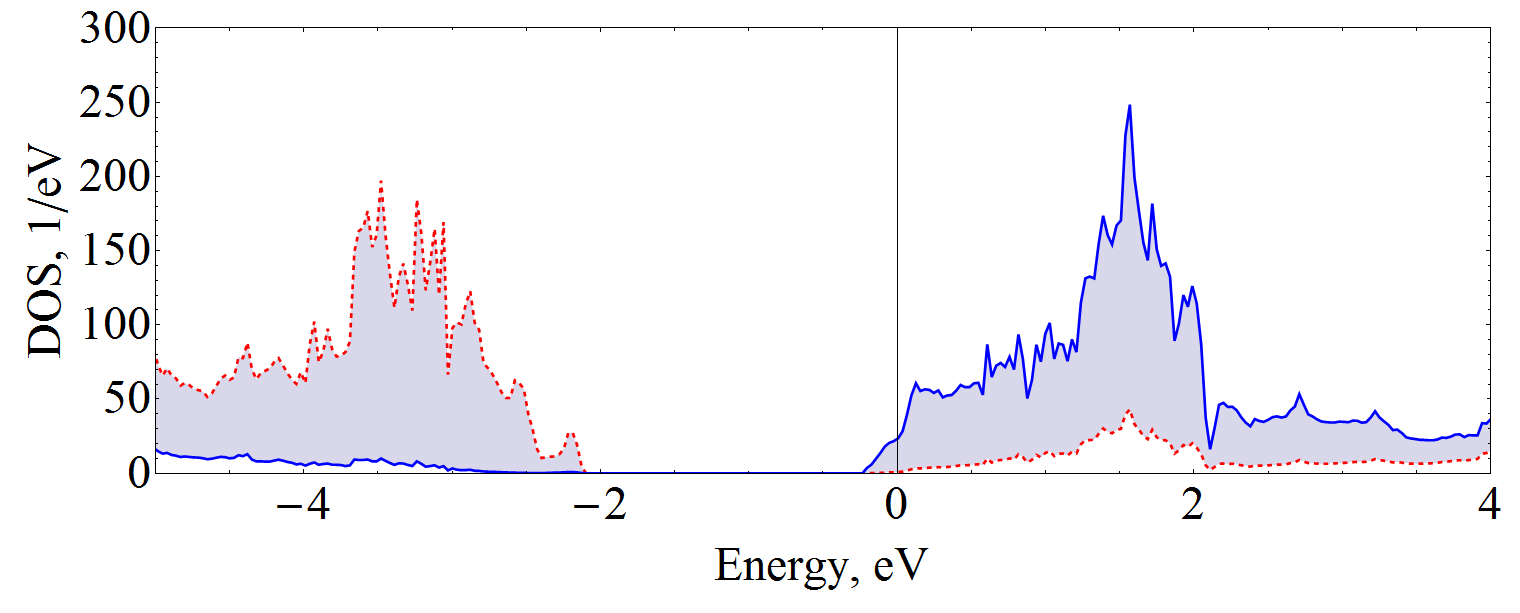}
           \label{fig:sto-dos-OandH} }
\caption{Left column: top views of \subref{fig:sto-dos-bare} bare 
                      $ 2 \times 2 $ $ {\rm SrTiO_3} $~surface as 
                      well as a surface with 
                      \subref{fig:sto-dos-1H} one hydrogen atom,  
                      \subref{fig:sto-dos-2H} two hydrogen atoms, 
                      \subref{fig:sto-dos-Ovac} an oxygen vacancy, 
                      and \subref{fig:sto-dos-OandH} an oxygen 
                      vacancy filled with a hydrogen atom. 
                      Strontium, titanium, oxygen, and 
                      hydrogen atoms are given in dark blue, light blue, 
                      red, and light grey, respectively, whereas 
                      oxygen vacancies are marked by a cross. 
         Right column: Corresponding partial electronic densities of 
                       states.}
\label{fig:sto-dos}
\end{figure}

The structures of the bare $ {\rm TiO_2} $-terminated $ {\rm SrTiO_3} $ 
slab  and the same slab with a hydrogen adatom are displayed in 
Fig.~\ref{fig:sto-charge}.  
Hydrogen atoms are adsorbed close to oxygen with bond lengths of 
0.986\,\AA\ and 0.982\,\AA, respectively, found for the $ 2 \times 1 $ 
and $ 2 \times 2 $ surface models in good agreement with previous 
observations. \cite{lin2009}  Top views and calculated densities of states 
for the bare slab and slabs with the same kinds of defects as studied 
for $ {\rm LaAlO_3} $ before are displayed in Fig.~\ref{fig:sto-dos}.  
In contrast to $ {\rm LaAlO_3} $, the bare $ {\rm SrTiO_3} $ slab as 
shown in Fig.~\ref{fig:sto-dos-bare} turns out to be an insulator with 
a band gap of about 1\,eV, much smaller than the value of about 2\,eV 
obtained for bulk $ {\rm SrTiO_3} $ with the same value of $ U $ for 
the Ti $ 3d $ orbitals.\cite{piyanzina2016,piyanzina2017} According to  additional 
calculations with varying numbers of layers in the $ {\rm SrTiO_3} $ 
slab we concluded that the minimum number of unit cells required to 
reproduce bulk electronic properties is four, which led us to use 
$ 4 \frac{1}{2} $ unit cells of $ {\rm SrTiO_3} $ in the central region 
of the $ {\rm LaAlO_3} $/$ {\rm SrTiO_3} $/$ {\rm LaAlO_3} $ 
heterostructure slab to be discussed in the following subsection. 

Eventually, the insulating nature of the $ {\rm SrTiO_3} $ slab can 
be traced back to the fact that, in contrast to $ {\rm LaAlO_3} $, 
within an ionic picture the alternating $ {\rm TiO_2} $ and 
$ {\rm SrO} $ layers are both neutral since all atomic shells can 
be filled within each layer separately. As a consequence, a \{001\} 
surface does not induce any missing charge or charge transfer and 
thus within the ionic picture the electronic state is the same as 
in the bulk material. 

Figs.~\ref{fig:sto-dos-1H} to \ref{fig:sto-dos-OandH} reveal the influence 
of defects on the structural and electronic properties of the slab. 
All kinds of defects taken into consideration cause atomic  
relaxations as well as a strong upshift of the Fermi energy with 
a finite occupation of the lowest conduction band states of one 
electron in the case of one hydrogen adatom (Fig.~\ref{fig:sto-dos-1H}) 
or an oxygen vacancy filled with hydrogen (Fig.~\ref{fig:sto-dos-OandH}) 
per $ 2 \times 2 $ supercell. In contrast, for two hydrogen adatoms 
(Fig.~\ref{fig:sto-dos-2H}) or a single oxygen vacancy 
(Fig.~\ref{fig:sto-dos-Ovac}) the conduction-band occupations rise to 
two. 

\begin{figure}[tb] %%%STO defects 
\vspace{4ex} 
\centering 
\subfigure{(a)
           \includegraphics[width=0.95\linewidth]{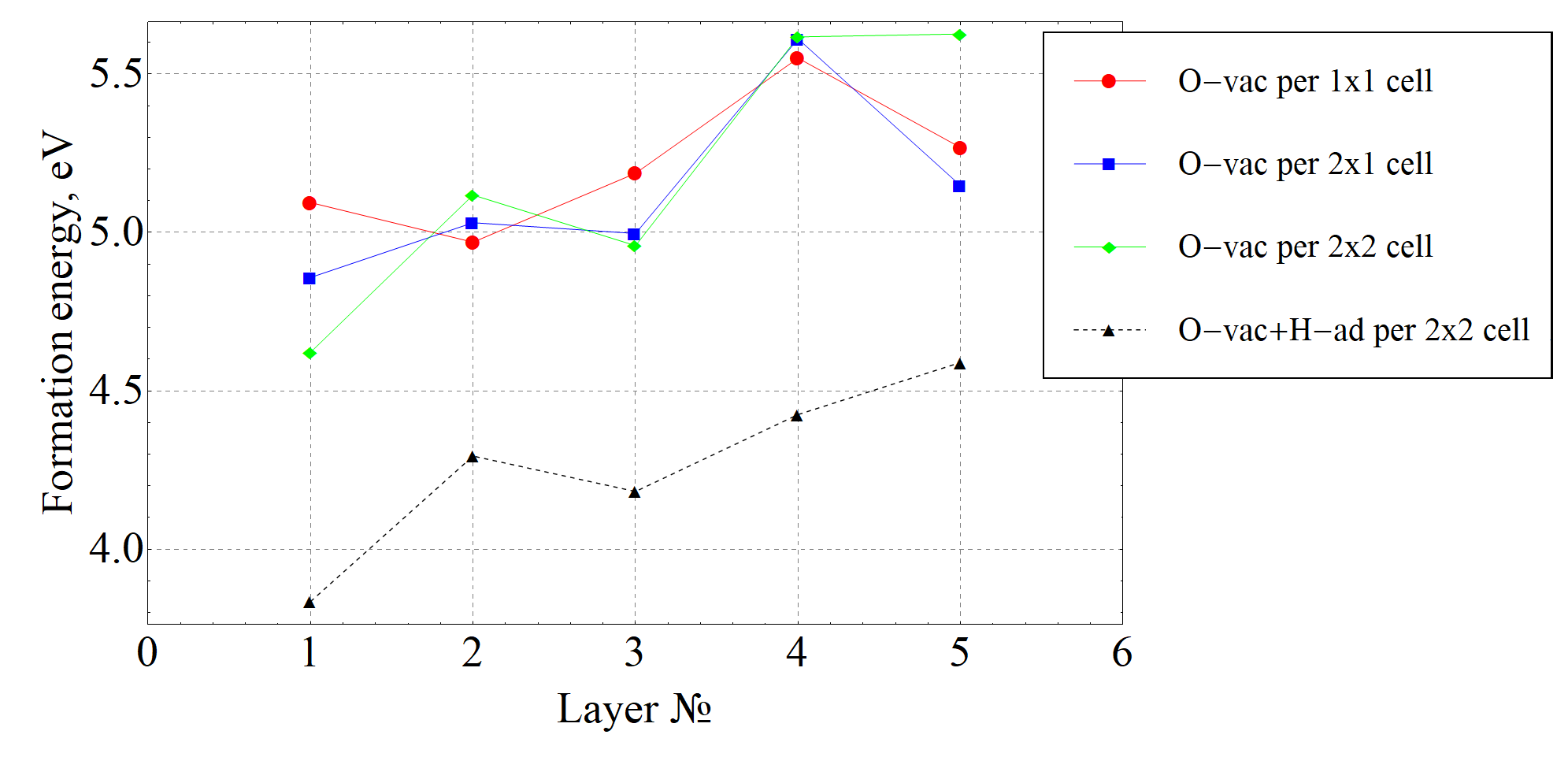} 
           \label{fig:sto-energy-O-vac-dep} } 
\subfigure{(b)
           \includegraphics[width=0.95\linewidth]{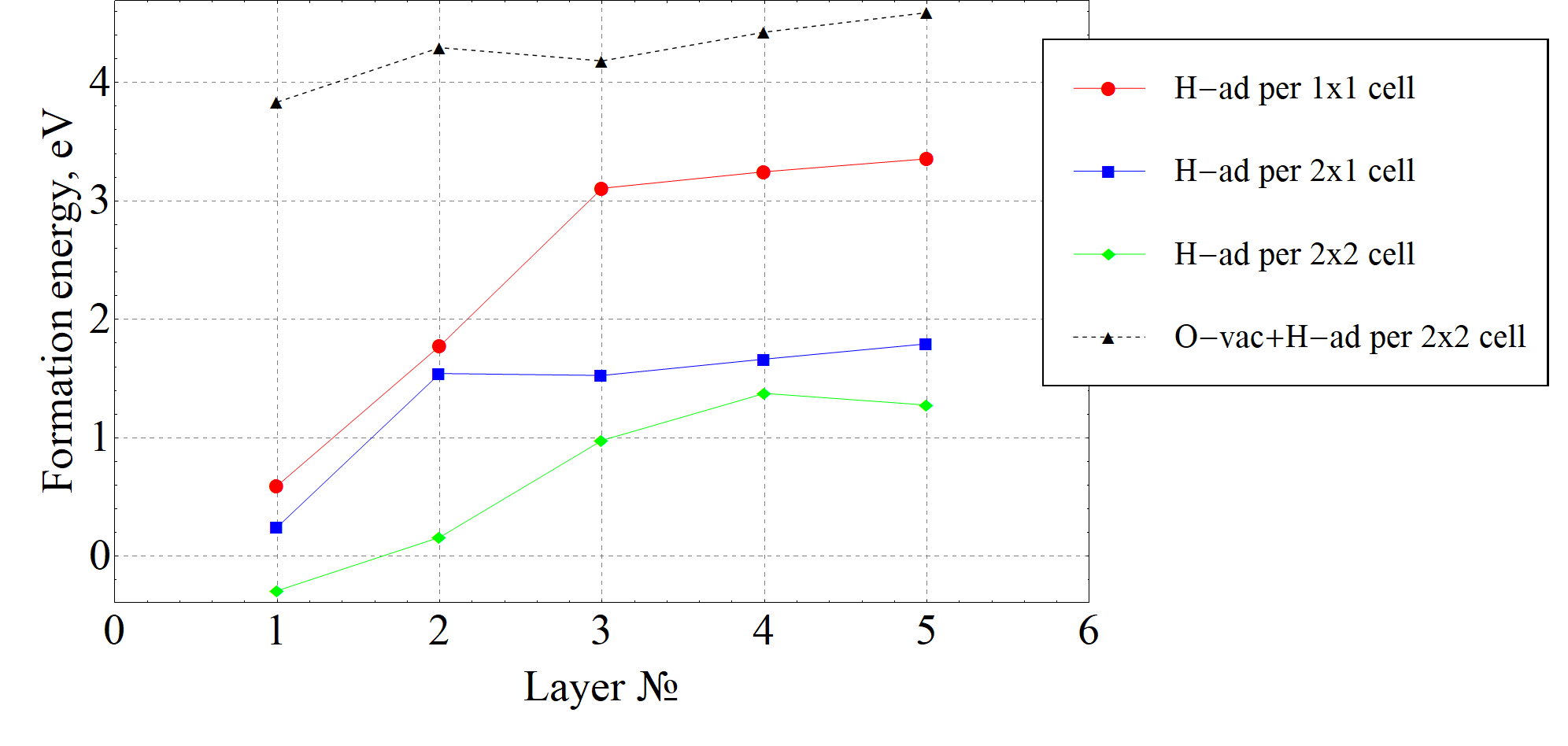} 
           \label{fig:sto-energy-H-ad-dep} }
\caption{Defect formation energies of $ {\rm TiO_2} $-terminated 
         $ {\rm SrTiO_3} $ slabs \subref{fig:sto-energy-O-vac-dep} with 
         an oxygen vacancy and \subref{fig:sto-energy-H-ad-dep} with a 
         hydrogen dopant atom in dependence of the position of the 
         defect. Layer 1 is the surface layer and counting 
         of layers treats SrO and $ {\rm TiO_2} $ layers separately.} 
\label{fig:sto-energy}
\end{figure}

Calculated energies of defect formation for different kinds of defects 
at various concentrations and at different positions are displayed in 
Fig.~\ref{fig:sto-energy}.  
As before, the curve for the case of an oxygen vacancy filled with 
a single hydrogen atom is included in both subfigures and serves as 
an orientation. For hydrogen dopant atoms the general trends are very 
similar to those already observed for the $ {\rm LaAlO_3} $ slabs. 
However, there is in general an upshift of all curves leaving only 
the case of a single surface hydrogen atom per $ 2 \times 2 $ surface 
unit cell with a negative energy of formation. 

The situation is much different for oxygen vacancies. Even though 
they show an overall increasing energy of formation at layers 
further away from the surface, the Coulomb repulsion between 
neighbouring vacancies seems to be considerably suppressed as compared 
to the situation in $ {\rm LaAlO_3} $ slabs. Specifically, for oxygen 
vacancies at the surface the difference in energy of defect formation 
between the lowest and highest concentration is 0.5\,eV in 
$ {\rm SrTiO_3} $, whereas it amounts to about 4.3\,eV in 
$ {\rm LaAlO_3} $. 
The tendency of increasing formation energy with layer depth and the
larger and smaller spreads in dependence on the layer was also observed
in Ref.~\onlinecite{silva2014} and is similar in our work.
It is not unreasonable to relate the reduction in the screening of the 
Coulomb repulsion between the charged defects to local correlations:
recently, it was observed that increasing $U$ on the La $4f$ orbitals of 
$ {\rm LaAlO_3}$/${\rm SrTiO_3}$ heterostructures reduces the lattice 
polarization within the LaO layers.~\cite{zabaleta2016,piyanzina2017} 
With this in mind one may expect the least screening of the Coulomb 
repulsion for the
layers with the strongest $U$. However such an argument is rather tentative
and needs further analysis which is beyond the scope of this work.

Finally, the values of defect formation energies for oxygen vacancies 
filled with a hydrogen dopant atom (dashed line in Fig.~\ref{fig:sto-energy}) 
lie between the formation energies of oxygen vacancies and hydrogen 
dopant atoms as expected and as also observed for LaAlO$_3$.

\subsection{$ {\bf LaAlO_3} $/$ {\bf SrTiO_3} $/$ {\bf LaAlO_3} $ slab}
\label{heterostructure}

A two-dimensional metallic electron system is formed at the interface 
when at least four unit cells of  $ {\rm LaAlO_3} $ (LAO) are grown on 
the $ {\rm TiO_2} $-terminated $ {\rm SrTiO_3} $ (STO) substrate,  
\cite{ohtomo2004,thiel2006}  which results in $ n $-type doping of the 
interface. Concomitant $ p $-type doping of the $ {\rm LaAlO_3} $ surface 
is expected from {\em ab initio} calculations 
\cite{lee2008,pentcheva2009} for the defect-free LAO-STO 
heterostructure on account of the polar catastrophe mechanism 
\cite{nakagawa2006} but has not been observed experimentally.  
\cite{thiel2006,berner2013a}

A promising explanation for the suppression of the surface metallic 
state  is a surface redox process, which leads to a release of bound 
electrons as free-carrier charges. \cite{bristowe2011,janotti2012} 
The most relevant example of such a process is oxygen vacancy formation. 
These vacancies are the most discussed type of defects in the literature 
from both experimental \cite{kalabukhov2007,park2013,salluzzo2013,yu2014} 
and theoretical \cite{zhong2010,li2011,zhang2010} perspectives. Moreover, 
oxygen vacancies are considered to be responsible for the formation of 
magnetic order. \cite{pavlenko2012b,pavlenko2013,lechermann2014} 

Usually, growth of $ {\rm LaAlO_3} $/$ {\rm SrTiO_3} $ heterostructures  
is complemented by a post-oxidation procedure at high pressures in order 
to reduce the amount of oxygen vacancies. \cite{scheiderer2015} Such 
a treatment is accompanied by surface contamination with water, 
hydrocarbons, and $ {\rm H}_{2} $ amongst others; samples could be also 
contaminated during experiments or storage. As was demonstrated in some 
theoretical studies, such adsorbates can donate electrons to the 
interface. \cite{son2010,li2013,yu2014} Bristowe {\em et al.}\ and
Scheiderer {\em et~al.}\ proposed a similar mechanism to explain 
additional charges due to atomic hydrogen adsorption, 
\cite{bristowe2011,scheiderer2015} which can be understood from the 
reaction 
\begin{displaymath}
{\rm H}^{0} + {\rm O}^{2-} \rightarrow  {\rm e}^{-}+ {\rm (OH)}^{-}    
               \;. 
\label{eqHad}
\end{displaymath}
The hydrogen atom bonds to the oxygen anion and therefore oxidizes under 
release of one electron per hydrogen atom. The electron moves to the 
interface resulting in an increase of the charge carrier density. In 
contrast, ionized hydrogen ($ {\rm H}^{+} $) cannot follow the same 
mechanism. The role of water and surface protonation was studied also 
from the viewpoint of conductivity switching. \cite{brown2016} It was 
shown that protonation can lead to a fully reversible change of 
conductance by more than four orders of magnitude. Water coating was 
also used for writing and erasing nanostructures at the interface. 
\cite{bi2012} 

\begin{figure}[t] %structures with charge transfer
\centering
\includegraphics[width=1.0\linewidth]{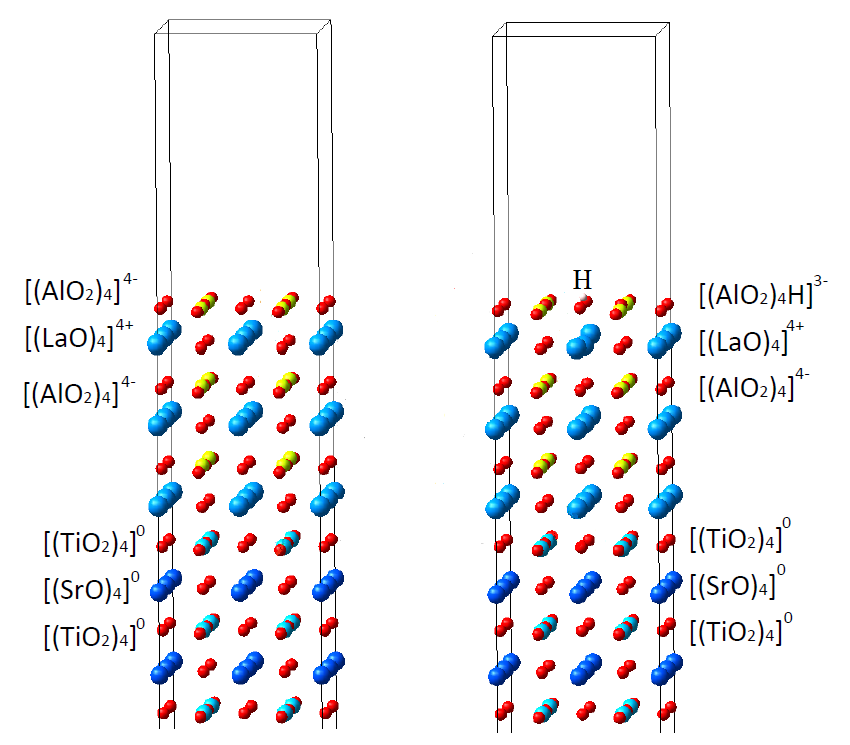}
\caption{Structures and ionic models of $ {\rm AlO_2} $-terminated 
         slabs of a $ {\rm LaAlO_3} $/$ {\rm SrTiO_3} $ heterostructure 
         with three LAO overlayers. Both a bare slab (left) and a slab 
         with one hydrogen adatom per $ 2 \times 2 $ surface unit cell 
         at the surface (right) are displayed. Only the upper part of 
         the slab cell is shown.}
\label{fig:het-structure}
\end{figure}

In the present work, we investigate the impact of defects (hydrogen 
dopants, oxygen vacancies as well as combination of these two) at various 
concentrations and various positions inside the $ {\rm LaAlO_3} $ 
and $ {\rm SrTiO_3} $ slabs on the structural and electronic properties. 
In particular, in the present section we demonstrate the influence of 
the aforementioned defects on the electronic properties of a 
$ {\rm LaAlO_3} $/$ {\rm SrTiO_3} $ heterostructure with three 
$ {\rm LaAlO_3} $ overlayers as displayed in Fig.\ \ref{fig:het-structure}. 

\begin{figure}[tb] %All DOSes
\centering
\subfigure{(a)
           \includegraphics[width=0.25\linewidth]{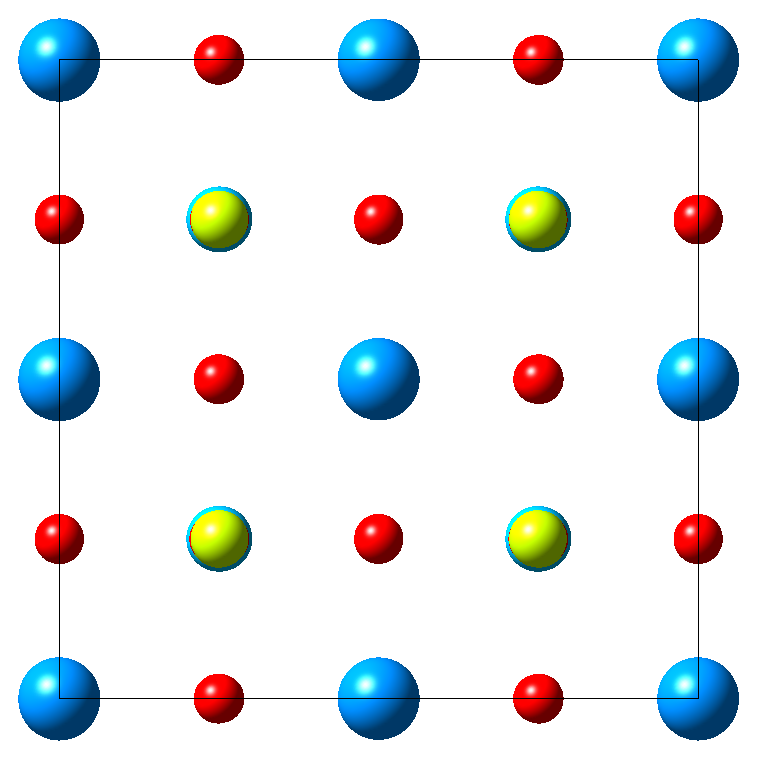}
           \includegraphics[width=0.68\linewidth]{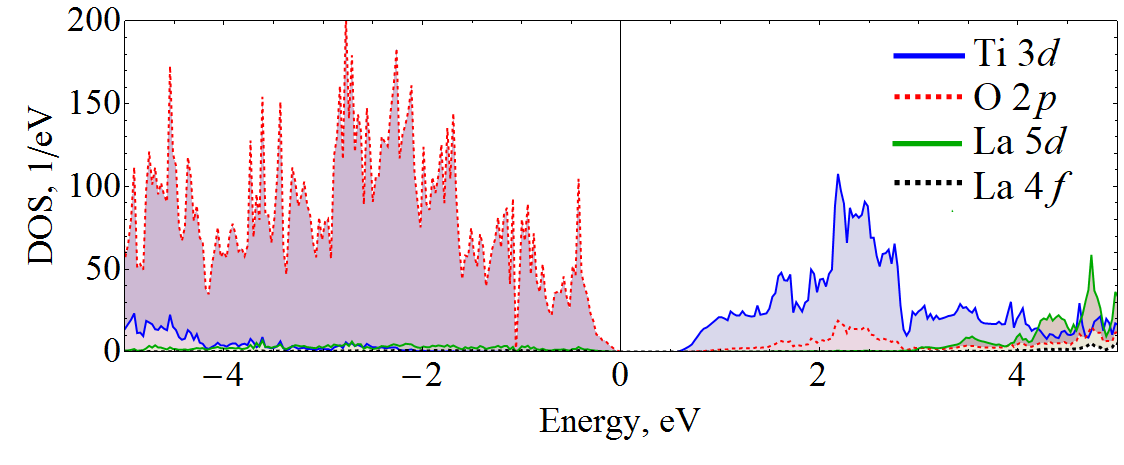}
           \label{fig:het-dos-bare} }\\
\subfigure{(b)
           \includegraphics[width=0.25\linewidth]{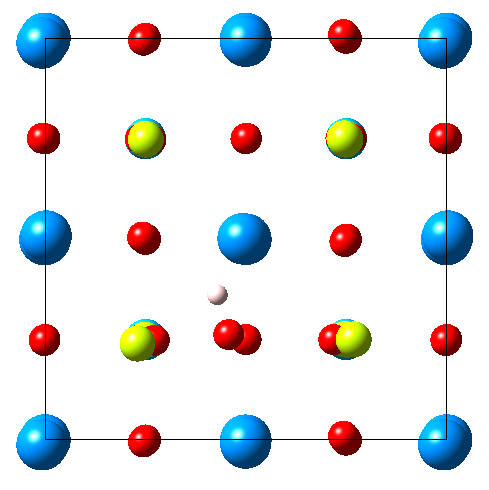}
           \includegraphics[width=0.68\linewidth]{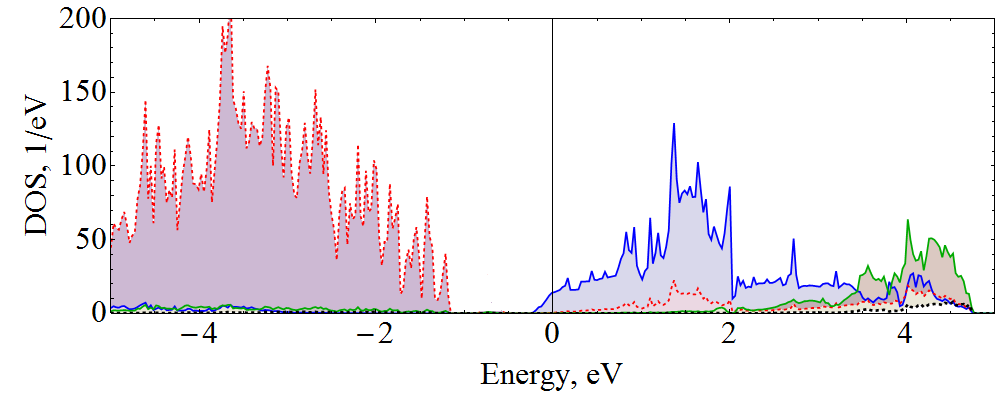}
           \label{fig:het-dos-1H} }\\
\subfigure{(c)
           \includegraphics[width=0.25\linewidth]{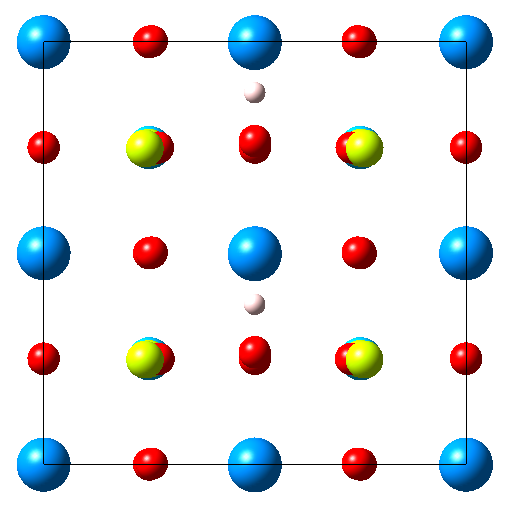}
           \includegraphics[width=0.68\linewidth]{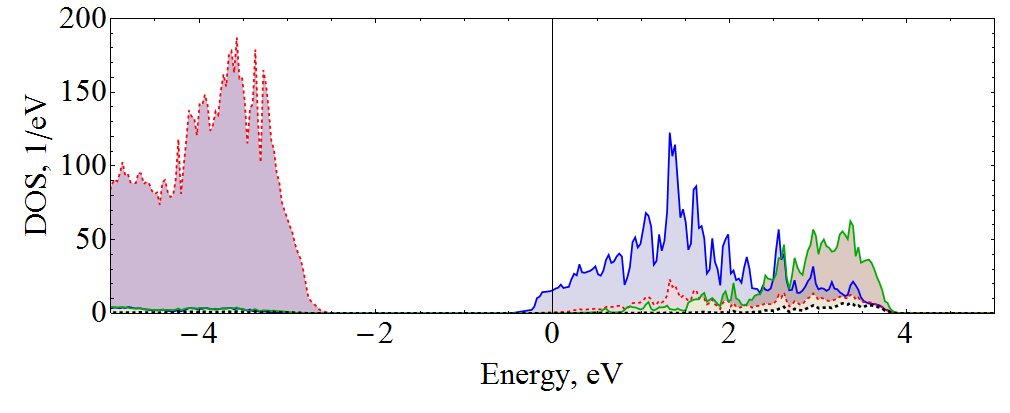}
           \label{fig:het-dos-2H}} \\
\subfigure{(d)
           \includegraphics[width=0.25\linewidth]{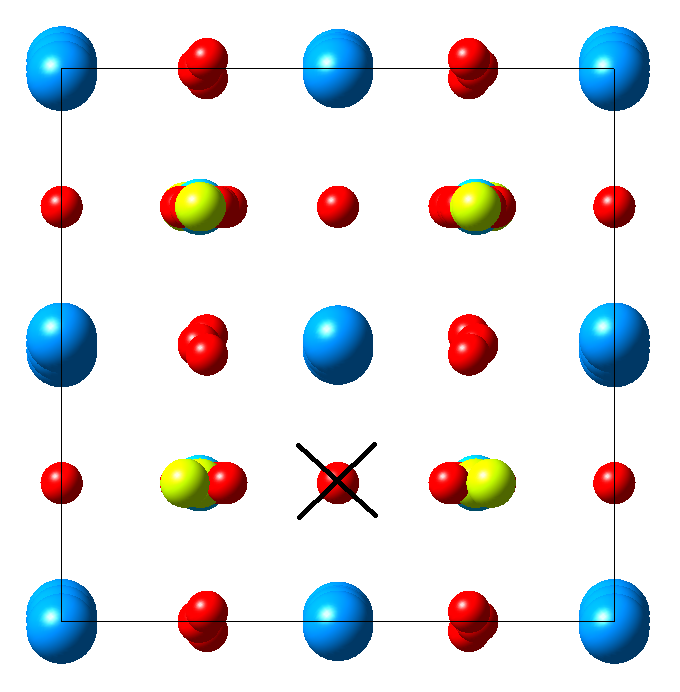}
           \includegraphics[width=0.68\linewidth]{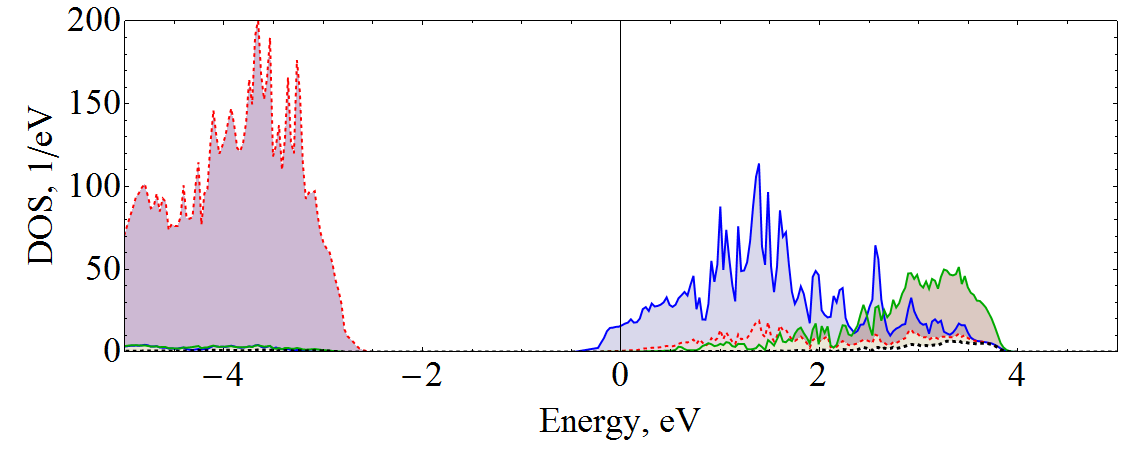}
           \label{fig:het-dos-Ovac} } \\
\subfigure{(e)
           \includegraphics[width=0.25\linewidth]{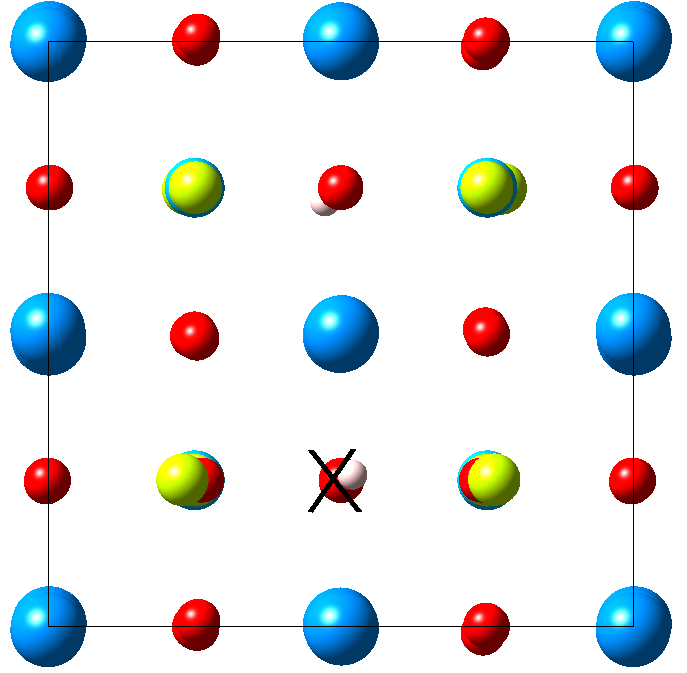}
           \includegraphics[width=0.68\linewidth]{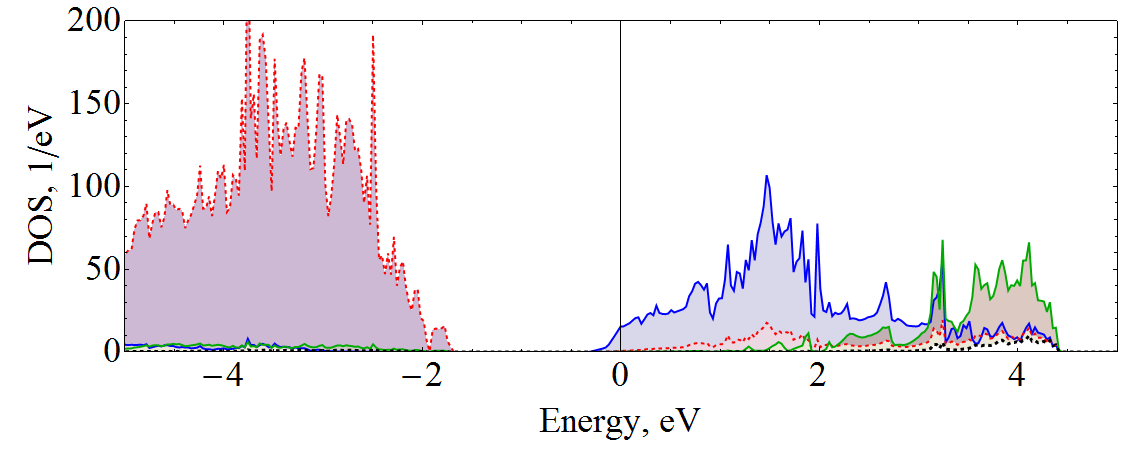}
           \label{fig:het-dos-OandH} }
\caption{Left column: top views of \subref{fig:het-dos-bare} bare 
                      $ 2 \times 2 $ $ {\rm LaAlO_3} $-surface of an 
                      $ {\rm LaAlO_3} $/$ {\rm SrTiO_3} $ heterostructure, 
                      \subref{fig:het-dos-1H} with one hydrogen atom on the 
                      surface, \subref{fig:het-dos-2H} with two hydrogen atoms, 
                      \subref{fig:het-dos-Ovac} with one oxygen vacancy, and 
                      \subref{fig:het-dos-OandH} with an oxygen vacancy filled 
                      with a hydrogen atom. Aluminum, lanthanum, oxygen, 
                      and hydrogen atoms are given in yellow, blue, red, 
                      and light grey, respectively, whereas oxygen 
                      vacancies are marked by a cross. 
         Right column: Corresponding partial electronic densities of 
                       states.}
\label{fig:het-dos}
\end{figure}

Crystal structures of the bare heterostructure and heterostructures 
with different kinds of defects and the corresponding calculated 
partial densities of states are displayed in Fig.\ \ref{fig:het-dos}.  
\begin{figure}[t] %Layer resolved
\centering
\includegraphics[width=0.87\linewidth]{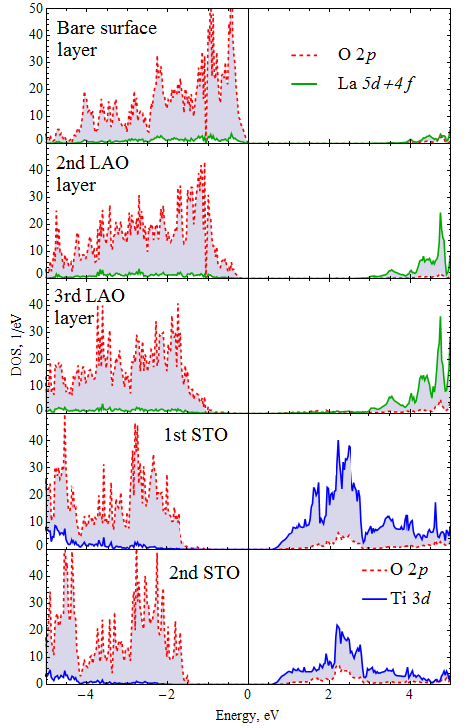}
\label{fig:het-bare-surface}
\caption{Layer-resolved DOS for a $ 2 \times 2 $ bare 
         $ {\rm LaAlO_3} $/$ {\rm SrTiO_3} $ heterostructure.}       
\label{fig:het-dos-layers-bare}
\end{figure}
Obviously, the bare heterostructure with three $ {\rm LaAlO_3} $ 
overlayers (Fig.\ \ref{fig:het-dos-bare}) has a band gap of about 
0.6\,eV. In contrast, all considered types of surface defects cause 
finite occupation of the Ti $ 3d $ states leading to metallic 
conductivity (see Figs.\ \ref{fig:het-dos-1H} to 
\ref{fig:het-dos-OandH}). These findings are not at all obvious 
from the ionic picture so successfully used above, within which  
bulk $ {\rm LaAlO_3} $ can be considered as a sequence of negatively 
charged $ {\rm (AlO_2)^{-}} $ and positively charged 
$ {\rm (LaO)^{+}} $ layers. That is, each LaO layer donates half 
an electron (per $ 1 \times 1 $ surface unit cell) to the neighboring 
$ {\rm AlO_2} $ layers on both sides. The situation is different for 
$ {\rm LaAlO_3} $ overlayers grown on top of $ {\rm SrTiO_3} $ (and 
for polar $ {\rm LaAlO_3} $ slabs with an equal number of LaO and 
$ {\rm AlO_2} $ layers). Still, according to the ionic picture each 
LaO layer would donate one electron and each $ {\rm AlO_2} $ layer 
would receive one electron (per layer unit cell) in order to avoid 
partially occupied atomic shells. However, in a slab geometry this 
would eventually signify a potential built-up from one side of the 
overlayer to the other. Since this situation is energetically unstable 
for sufficiently thick $ {\rm LaAlO_3} $ overlayers, the oxygen 
states at the surface are not completely filled whereas the La $ 5d $ 
states at the interface are expected to be partially filled and 
metallic conductivity would arise in these two layers. In other words, 
both surfaces of the overlayer, i.e.\ both the interface with the 
neighboring $ {\rm TiO_2} $ layer and the $ {\rm AlO_2} $-terminated 
surface layer would become metallic whereby the charge carriers at 
the interface in fact reside in the Ti $3d$ states, not in the 
La $5d$ states. Again, this line of reasoning is the same for a 
polar $ {\rm LaAlO_3} $ slab. While this scenario is indeed obtained 
from calculations for unrelaxed structures, structural relaxation 
induces dipole fields in the LAO planes,\cite{pentcheva2009,pavlenko2011} 
which counteract the field from the polar setup, and thereby restore 
semiconducting behavior for LAO slabs below a critical thickness, as 
is indeed observed in Fig.~\ref{fig:het-dos-bare} for the case of 
three $ {\rm LaAlO_3} $ overlayers. 

Starting from the semiconducting state observed for the bare 
heterostructure with three overlayers, the defects considered here 
all introduce metallic conductivity at the interface. This may be 
induced by either two H-adatoms per $ 2 \times 2 $ surface cell, 
see Fig.~\ref{fig:het-dos-2H}, or by an oxygen vacancy within the 
same surface unit cell, see Fig.~\ref{fig:het-dos-Ovac}. Both 
situations lead to a surface layer with formal charge $ -4 $ and 
occupation of the lowest part of the conduction band, which is of 
Ti $ 3d $ character. 

\begin{figure}[t] %Layer resolved
\centering
\includegraphics[width=0.8\linewidth]{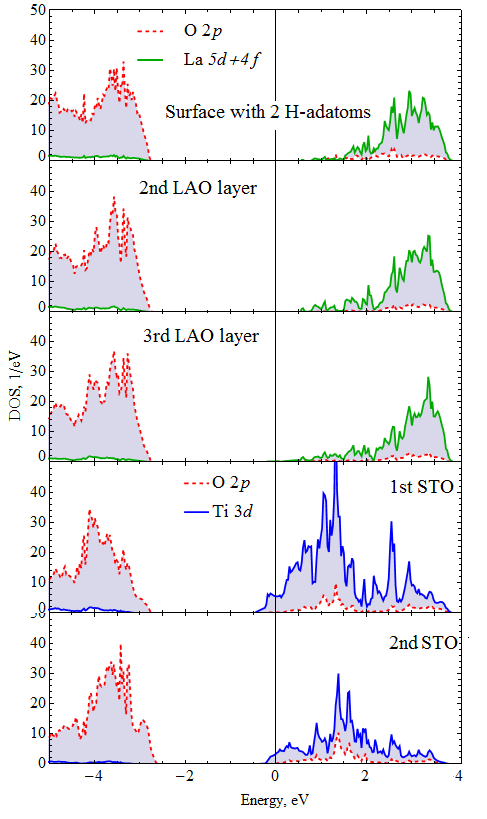}
\label{fig:het-withHatsurface}
\caption{Layer-resolved DOS for a $ 2 \times 2 $ 
         $ {\rm LaAlO_3} $/$ {\rm SrTiO_3} $ heterostructure       
         with two hydrogen adatoms at its surface.} 
\label{fig:het-dos-layers-2H}
\end{figure}

\begin{figure}[t]  % defects vs layer
\centering 
\includegraphics[width=0.9\linewidth]{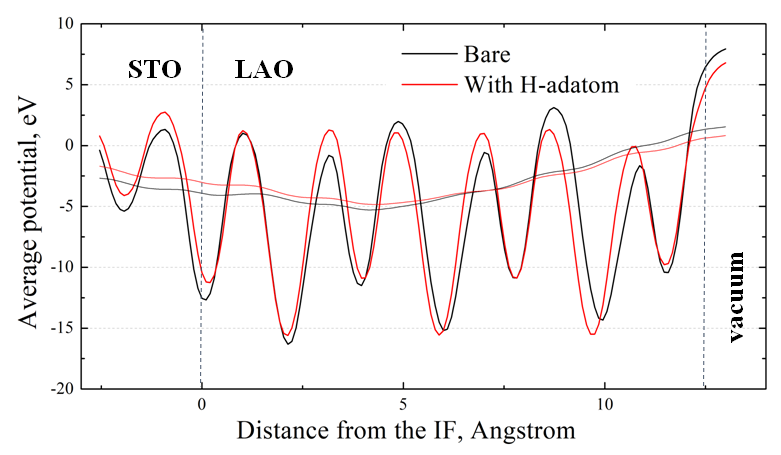} 
\caption{Average effective single-particle potential of 
         $ {\rm AlO_2} $-terminated $ {\rm LaAlO_3} $/$ {\rm SrTiO_3} $ 
         heterostructures with three $ {\rm LaAlO_3} $ overlayers. The 
         black and red curve refer to the bare heterostructure and 
         the heterostructure with two hydrogen surface adatoms per 
         $ 2 \times 2 $ surface cell, respectively, corresponding to 
         the situations shown in Figs.\ \protect 
         \ref{fig:het-dos-layers-bare} and \ref{fig:het-dos-layers-2H}.
         The smooth curves were generated from the original ones by 
         proper broadening to allow for a better comparison of the 
         overall trends.}
\label{fig:het-potential} 
\end{figure}

It is very instructive to study the effect of hydrogen adatoms on the 
electronic structure by means of the layer-resolved densities of 
states. The corresponding results for the bare heterostructure and the 
heterostructure with two hydrogen atoms at the surface are shown in 
Figs.~\ref{fig:het-dos-layers-bare} 
and \ref{fig:het-dos-layers-2H}, respectively. 
Note that these layer-resolved densities of states correspond to the 
densities of states displayed in Figs.~\ref{fig:het-dos-bare} and 
\ref{fig:het-dos-2H}, respectively. In Fig.~\ref{fig:het-dos-layers-bare} 
we clearly observe a downshift of the electronic states on going from 
the surface to the interface, which we attribute to the built-up of 
electric potential mentioned above. Indeed, this potential shift is 
clearly observed in Fig.~\ref{fig:het-potential},  
which displays the effective single-particle potential across the slab 
averaged over planes parallel to the surface. While for the bare 
heterostructure all layers are semiconducting as expected from the 
densities of states shown in Fig.~\ref{fig:het-dos-bare}, the metallic 
behavior of the heterostructure with hydrogen adatoms results from  
filling of the Ti $ 3d $ states at the interface, since the electrons 
provided by the H adatoms make charge transfer from the interface to 
the surface as observed for the bare heterostructure obsolete. At the 
interface the carriers are then attracted by the $ {\rm TiO_2} $ 
layer due to the higher electronegativity of Ti as compared to La and 
this causes the metallic conductivity in this layer. As a consequence 
of the reduced charge transfer across the LAO overlayers, the potential 
difference between surface and interfaces is considerably reduced as 
is indeed observed in Fig.~\ref{fig:het-potential} as well as in 
Fig.~\ref{fig:het-dos-layers-2H}, where the valence band maximum 
formed from the O $ 2p $ states in the LAO overlayer stays at the 
same energetical position throughout the slab and also the lower edge 
of the La $ 5d $ states barely shifts. 

The just outlined passivation of the surface by adsorption of hydrogen, 
which is equivalent to the situation of a surface with oxygen vacancies 
as sketched in Fig.~\ref{fig:het-dos-Ovac} and in qualitative agreement 
with Refs.~\onlinecite{bristowe2011,janotti2012} thus naturally explains 
why no built-up of electric potential has been observed in experiments 
\cite{segal2009,slooten2013,berner2013b} and at the same time the surface 
is insulating. \cite{thiel2006,berner2013b} However this does not yet 
explain the observation of a critical thickness of the LAO overlayer 
necessary to form a conducting interface, which we will address below. 

Finally, for the case of one excess electron provided by one hydrogen adatom 
per $ 2 \times 2 $ surface cell as presented in Fig.~\ref{fig:het-dos-1H} 
and the equivalent situation of an oxygen vacancy filled by a hydrogen 
atom as displayed in Fig.~\ref{fig:het-dos-OandH} surface passivation 
is not complete and the just discussed effects become less pronounced 
but still cause metallic conductivity at the interface with a smaller 
filling of the Ti $ 3d $ states.

\begin{figure}[tb]  % defects vs layer
\vspace{4ex} 
\centering 
\subfigure{(a)
           \includegraphics[width=0.95\linewidth]{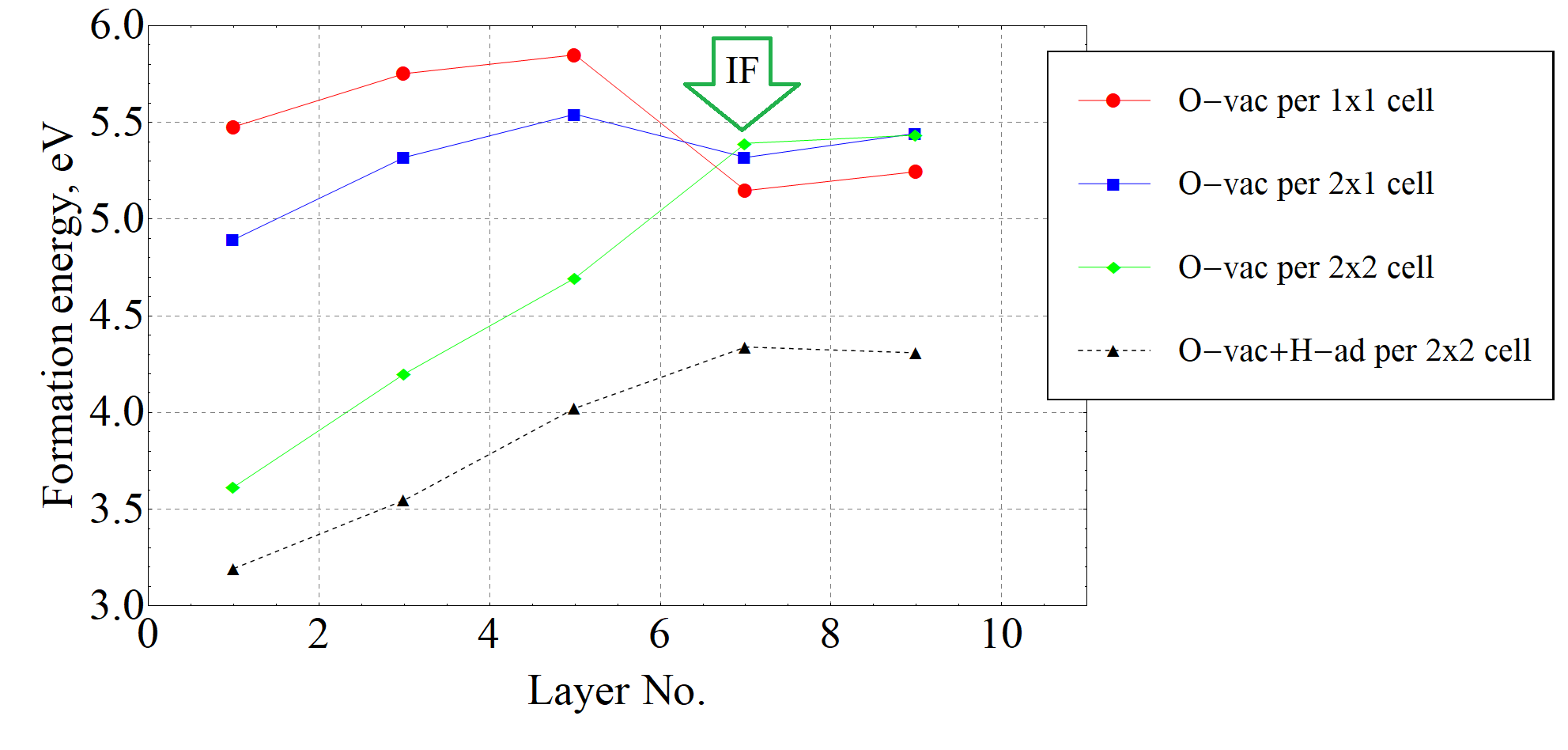} 
           \label{fig:het-energy-O-vac-dep} }  
\subfigure{(b)
           \includegraphics[width=0.95\linewidth]{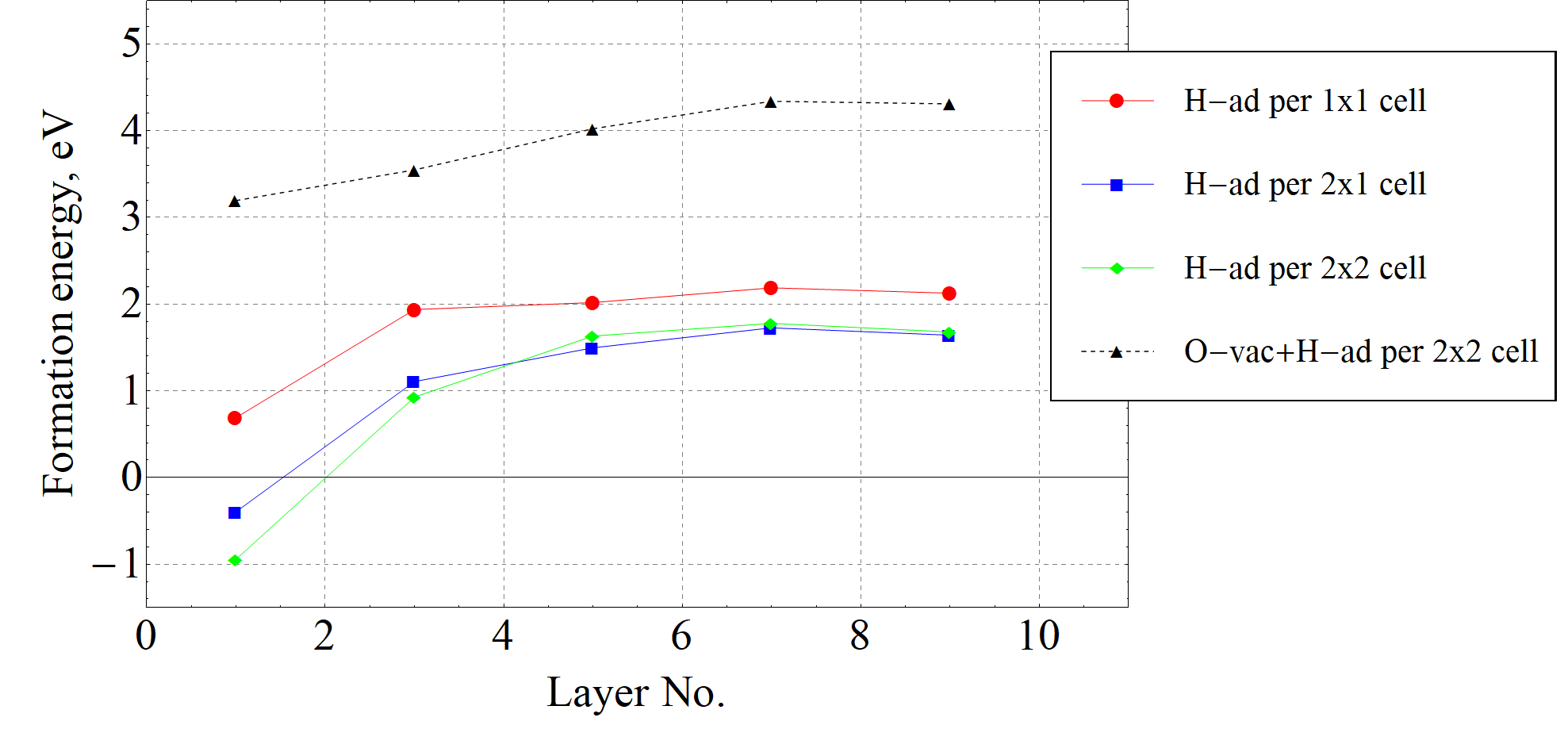} 
           \label{fig:het-energy-H-ad-dep} }
\caption{Defect formation energies of $ {\rm AlO_2} $-terminated 
         $ {\rm LaAlO_3} $/$ {\rm SrTiO_3} $ heterostructures 
         \subref{fig:het-energy-O-vac-dep} with an oxygen vacancy 
         and \subref{fig:het-energy-H-ad-dep} with a hydrogen 
         adatom in dependence of the position of the defect. 
         Layer 1 is the surface layer and counting of layers treats 
         LaO, $ {\rm AlO_2} $, SrO and $ {\rm TiO_2} $ layers separately.} 
\label{fig:het-energy}
\end{figure}

The formation energies calculated for various types of defects at 
various concentrations are displayed in Fig.~\ref{fig:het-energy}.  
Here we also account for buried point defects by placing them in layers 
below the surface extending even into the STO substrate. For both hydrogen 
dopants and oxygen vacancies we observe increase of the formation energy 
with increasing concentration as a consequence of their Coulomb repulsion. 
In addition, formation energies generally are lowest for defects at the 
surface. However, especially formation energies of hydrogen dopants 
saturate beyond the second layer due to effective screening. This is 
different for oxygen vacancies, which show strong increase of the 
formation energies towards deeper LAO layers and show saturation only 
within the STO substrate. For higher concentrations this behavior leads 
even to a strong decrease of the formation energy between the fifth 
and seventh layer (counted from the surface) as was also observed for 
four $ {\rm LaAlO_3} $ overlayers. \cite{pavlenko2012b} Nevertheless, 
according to these results only hydrogen adatoms at rather low 
concentration will form stable defect configurations, whereas all other 
defects show positive formation energies.  

In passing we point to the value of about 0.7 eV found for the formation 
energy of a hydrogen adatom at the surface of a $ 1 \times 1 $ surface 
cell, which agrees rather well with the value of about 0.5 eV given by 
Son and coworkers.\cite{son2010} 

\begin{figure}[tb]  % defects vs layer
\centering 
\includegraphics[width=0.7\linewidth]{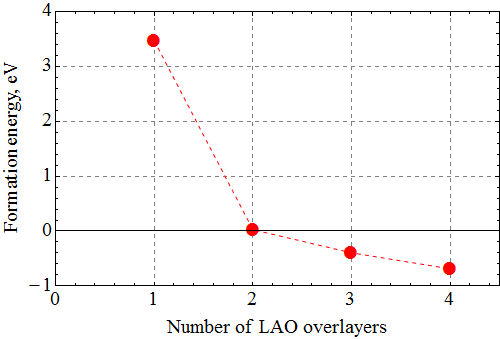} 
\caption{Defect formation energies of $ {\rm AlO_2} $-terminated 
         $ {\rm LaAlO_3} $/$ {\rm SrTiO_3} $ heterostructures with 
         one H adatom per $ 2 \times 1 $ surface cell as a function 
         of the number of LAO overlayers. Note that the data point 
         for 3 LAO overlayers is identical to the blue data point 
         for layer~1 in Fig.~\ref{fig:het-energy}.}
\label{fig:het-energy-thick}
\end{figure}

As mentioned before, the surface passivation scenario sketched above, 
while nicely explaining the suppression of the shift of both the effective 
potential and the band edges across the LAO overlayer as well as the 
semiconducting and metallic behavior of the surface and interface, 
respectively, does not yet provide an explanation for the onset of 
the latter situation observed at four LAO layers and beyond. In this 
context it is very helpful to consider the formation energy of hydrogen  
adatoms as a function of the number of LAO overlayers, which is 
displayed in Fig.~\ref{fig:het-energy-thick}. 
Note that the value of the defect formation energy for three LAO 
overlayers shown in this figure is identical to the value given 
for layer 1 for one hydrogen per $ 2 \times 1 $ surface cell 
given in Fig.\ \ref{fig:het-energy}. Obviously, stable adsorption of 
hydrogen is unlikely for thin LAO overlayers and only at a critical 
thickness of the LAO slab do H adatoms form a stable connection 
with the surface, which eventually causes the metallic interface 
as outlined above. For the investigated set-up this critical thickness 
is slightly above 2 LAO overlayers whereas experiments rather suggest 
a critical thickness between 3 and 4 LAO layers.\cite{thiel2006} 
However, the tendency to bind hydrogen at the surface of this 
heterostructure with increasing thickness of the LAO slab is clearly 
demonstrated in Fig.~\ref{fig:het-energy-thick}. Furthermore, note 
that according to Fig.\ \ref{fig:het-energy} the formation energies 
depend somewhat on the defect concentration. We thus take this finding 
as a proof of principle for a critical thickness which eventually 
controls the transition to a metallic interface state through binding 
of surface adatoms. 
Similar conclusions have been drawn in
Refs.~\onlinecite{bristowe2011,janotti2012,bristowe2014,krishnaswamy2015},
and our results qualitatively agree with those of Ref.~\onlinecite{son2010} 
though there a  $ 1 \times 1 $ surface cell is taken.

\section{Impact of impurities on magnetic properties}
\label{magnetism}

As mentioned in the introduction, the origin of the magnetic order 
forming at the interface below room temperature,\cite{brinkman2007} 
which coexists with a superconducting state below 300~mK, 
\cite{luli2011a,moler2011} is still a matter of discussion. 
\cite{pavlenko2012a,pavlenko2012b,michaeli2012,kalisky2012,pavlenko2013,lechermann2014,ruhman2014,yu2014} 
However, from {\it ab initio} calculations including either static 
($ +U $)~\cite{pavlenko2012a,pavlenko2012b} or dynamic 
(+DMFT)~\cite{lechermann2014} correlations it was concluded that 
ferromagnetic order can be induced by oxygen vacancies, whereas the bare 
heterostructure would not show a magnetic instability. In particular, 
oxygen vacancies at the interface lead to an orbital reconstruction 
at adjacent Ti sites and generate a local spin polarization. 
Schemes to extend the analysis of local moment formation to intermediate 
and low oxygen vacancy concentrations have been investigated in 
Refs.~\onlinecite{pavlenko2013} and \onlinecite{behrmann2015}. 

While the previous studies dealt with heterostructures with four 
LAO overlayers, here we focus on 3LAO/STO heterostructures and 
consider the role of oxygen vacancies in the $ {\rm AlO_2} $ 
surface plane and in the $ {\rm TiO_2} $ interface plane for the 
onset of ferromagnetism by performing spin-polarized calculations. 
In addition, we investigate, for the first time, the impact of 
hydrogen dopants on local-moment formation.

\subsection{Oxygen vacancy}
\label{mag-oxygen}

\begin{figure}[tb] %%O-vac DOSes
\subfigure{(a)
           \includegraphics[width=0.25\linewidth]{hetero1}
           \includegraphics[width=0.68\linewidth]{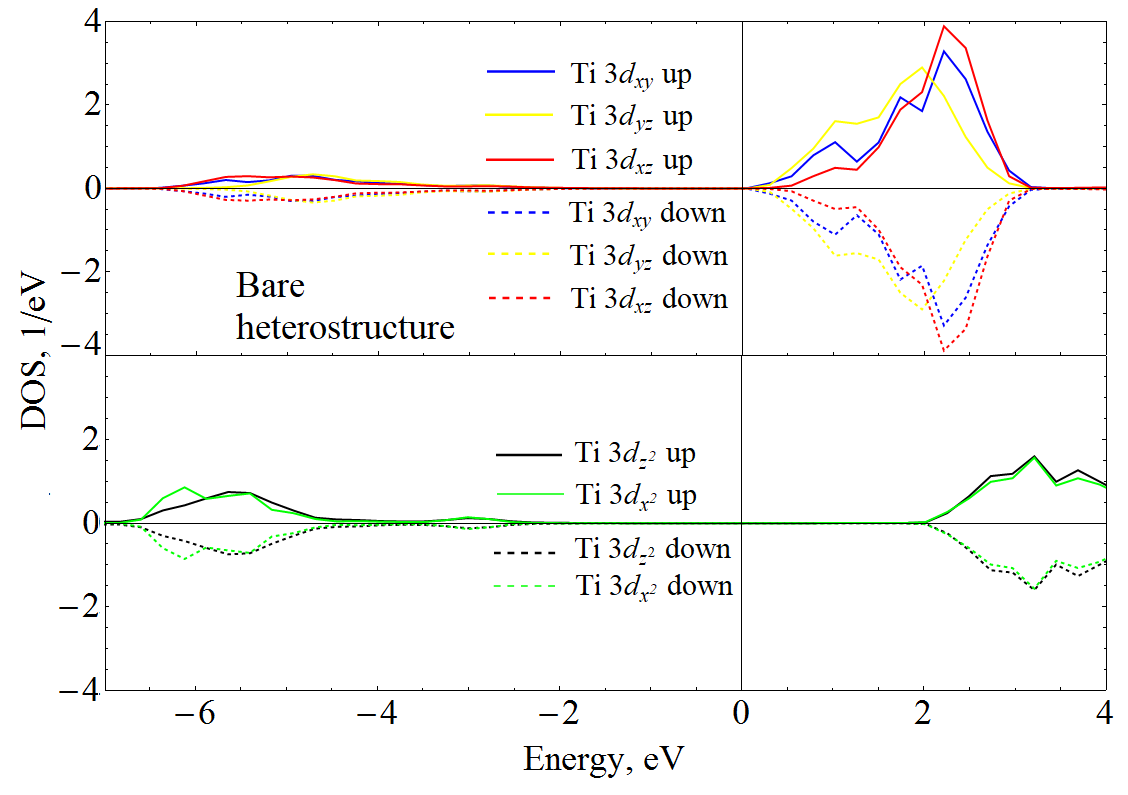}
           \label{fig:het-mag-dos-bareTi} }
\subfigure{(b)
           \includegraphics[width=0.25\linewidth]{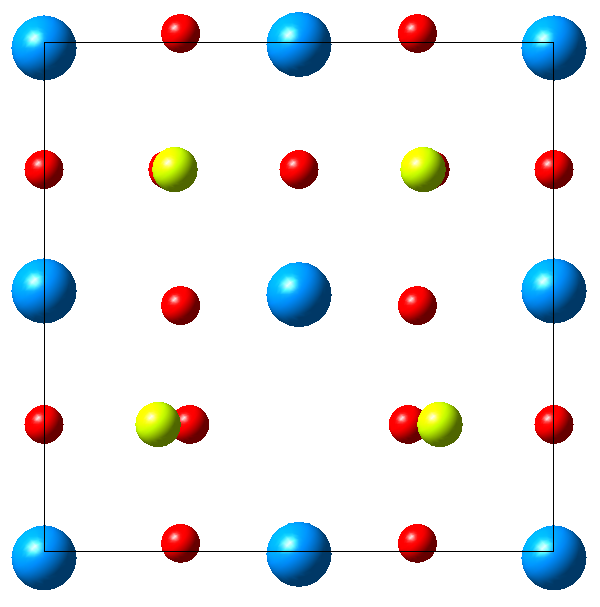}         
           \includegraphics[width=0.68\linewidth]{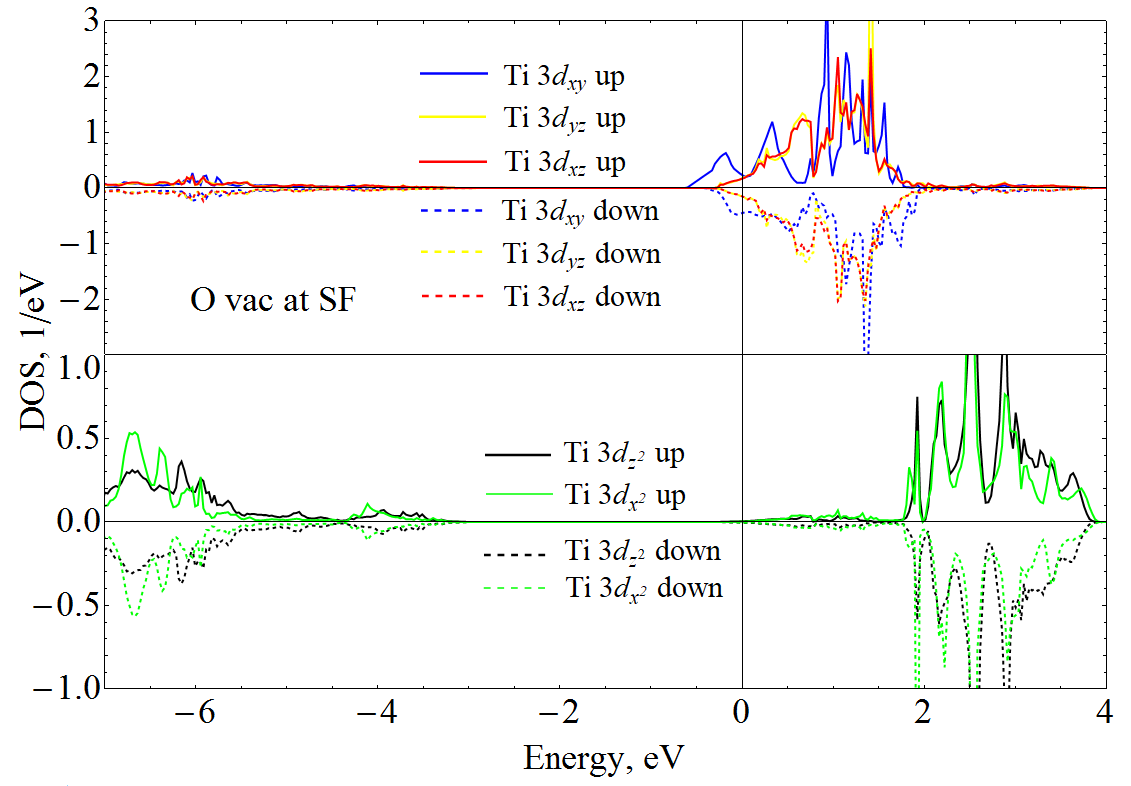} 
           \label{fig:het-mag-dos-OvacTi} }
\subfigure{(c)
           \includegraphics[width=0.25\linewidth]{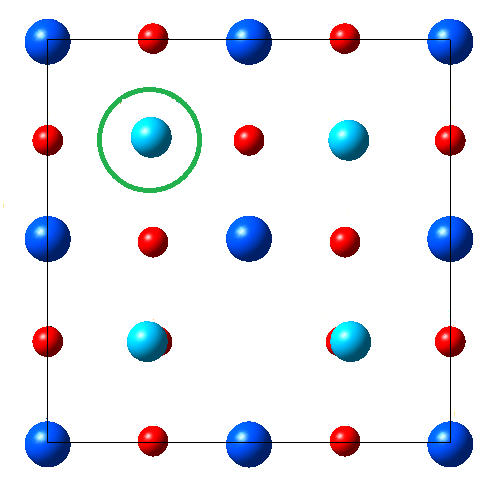}
           \includegraphics[width=0.68\linewidth]{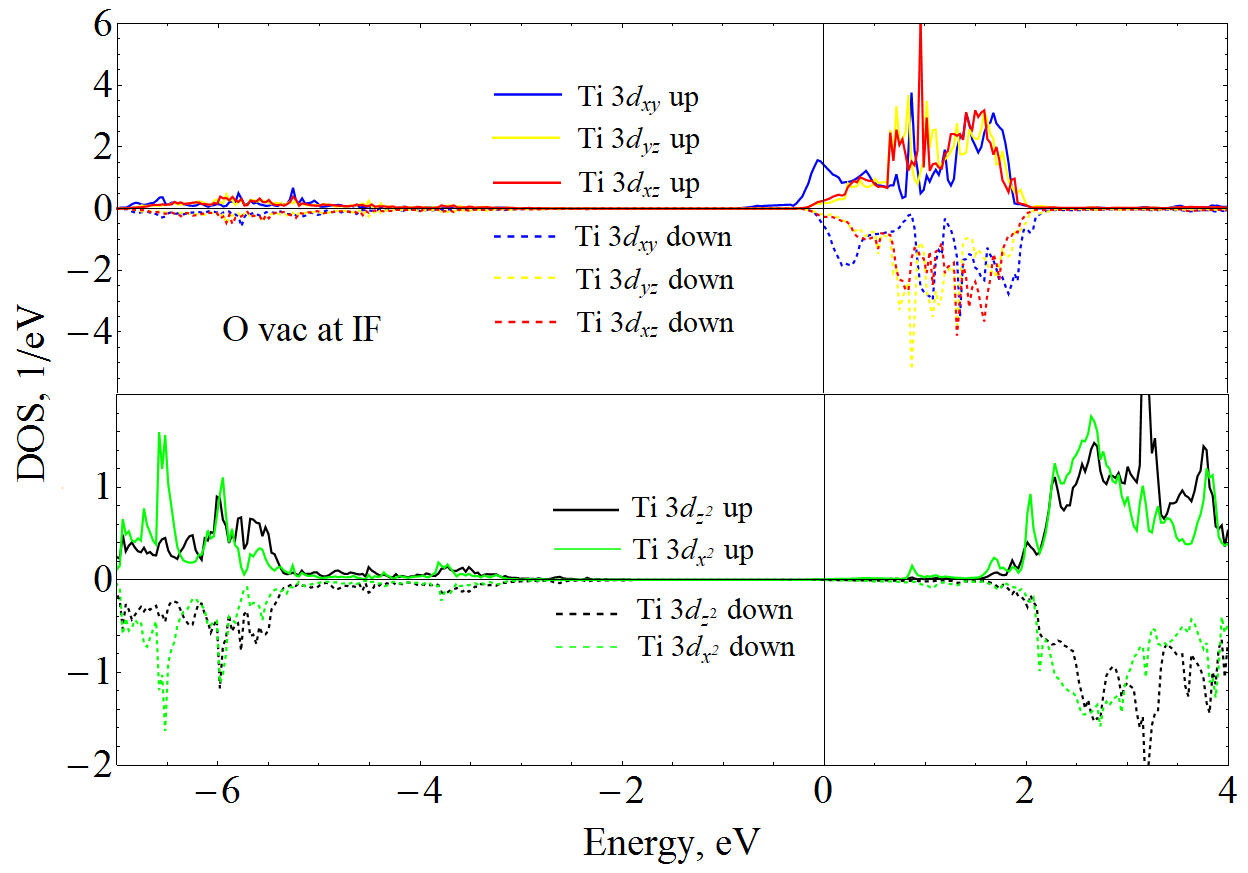}
           \label{fig:het-mag-dos-Ofar} }
\subfigure{(d)
           \includegraphics[width=0.25\linewidth]{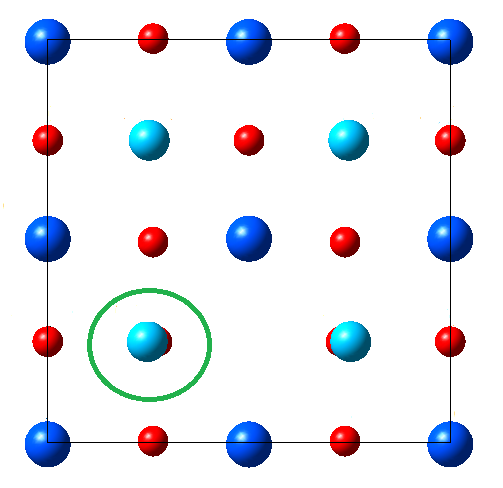}
           \includegraphics[width=0.68\linewidth]{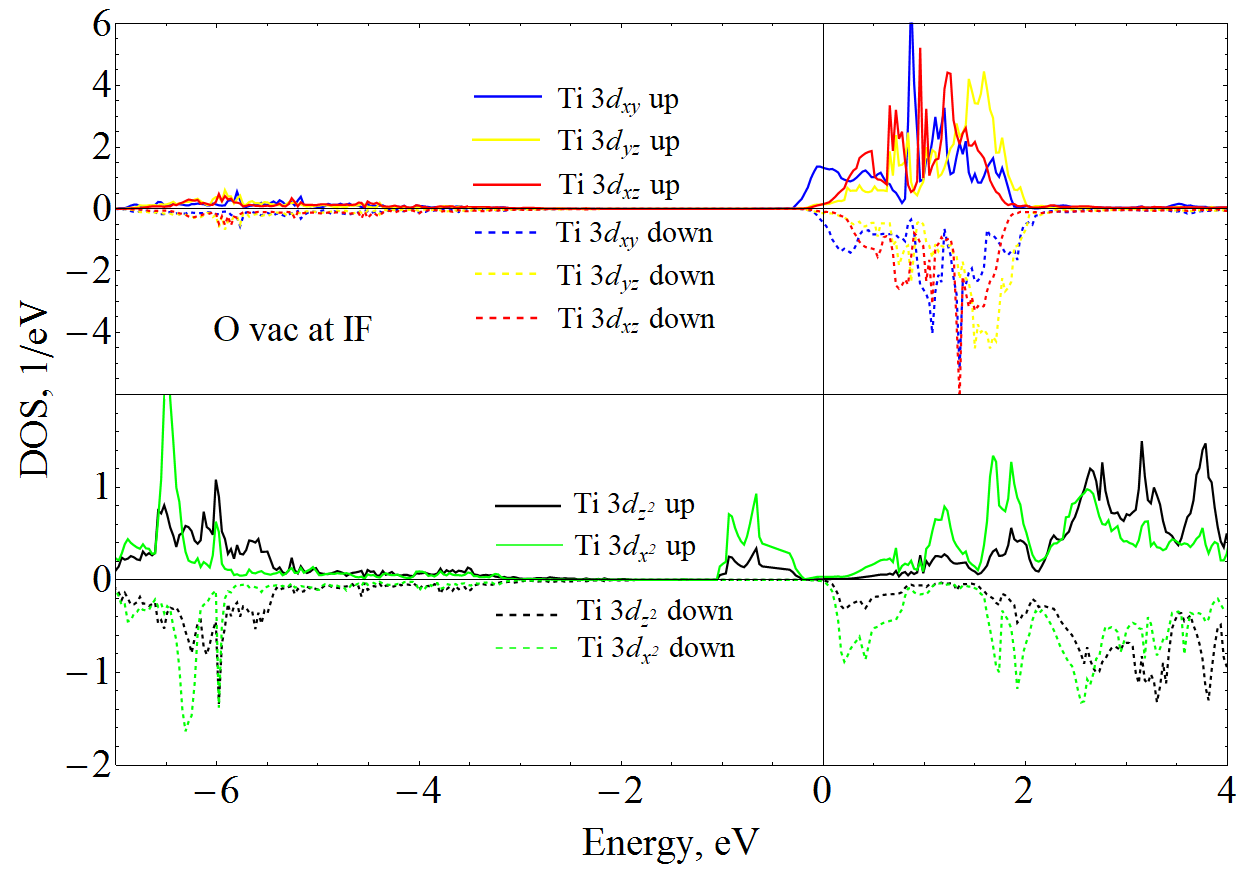}
           \label{fig:het-mag-dos-Onear} }
\caption{Left column: top views of 
         \subref{fig:het-mag-dos-bareTi} bare $ 2 \times 2 $ 
               $ {\rm LaAlO_3} $-surface of an 
               $ {\rm LaAlO_3} $/$ {\rm SrTiO_3} $ heterostructure 
               and with oxygen vacancy located in 
         \subref{fig:het-mag-dos-OvacTi} 
               the surface AlO$ _{2} $ layer or 
         \subref{fig:het-mag-dos-Ofar}/\subref{fig:het-mag-dos-Onear} 
               the interfacial TiO$_{2}$ layer. 
         Aluminum, lanthanum, oxygen,  strontium, titanium, and 
         hydrogen atoms are given in yellow, blue, red,  dark blue, 
         light blue, and light grey, respectively, whereas oxygen 
         vacancies are marked by a cross. 
         Right column: Corresponding spin- and orbital-projected 
         Ti $ 3d $ electronic densities of states of 
         \subref{fig:het-mag-dos-bareTi} all Ti atoms, 
         \subref{fig:het-mag-dos-OvacTi} an interfacial Ti atom, 
         \subref{fig:het-mag-dos-Ofar}  an interfacial Ti atom distant from 
                the O vacancy and 
         \subref{fig:het-mag-dos-Onear} an interfacial Ti atom close to 
                the O vacancy.}  
\label{fig:het-mag-dos-Ovac}
\end{figure}

Being motivated by the nontrivial finding that for sufficiently large 
concentrations of oxygen vacancies the formation energy drops at the 
interface (see Fig.~\ref{fig:het-energy-O-vac-dep}) we reexplored the 
distinct orbital contributions of Ti atoms to the spin-resolved 
density of states. In doing so, we focused especially on two different 
scenarios, namely, in addition to the bare heterostructure, 
heterostructures with oxygen vacancies either in the surface 
$ {\rm AlO_2} $ layer or in the $ {\rm TiO_2} $ layer at the interface. 
The results of the corresponding spin-polarized calculations are 
displayed in Fig.\ \ref{fig:het-mag-dos-Ovac}. 
While the spin- and orbital-resolved $ 3d $ densities of states of 
all Ti atoms of the bare 3LAO/STO heterostructure as shown in 
Fig.~\ref{fig:het-mag-dos-bareTi} clearly reveal the insulating 
behaviour and zero magnetic moment of this situation, the presence 
of an oxygen vacancy at the surface causes an upshift of the Fermi 
energy and thus finite conductivity as has been already discussed 
in connection with Fig.~\ref{fig:het-dos-Ovac}. Here, we obtain, in 
addition, a finite spin-splitting especially of the Ti $ 3d $ $ t_{2g} $ 
states at the interface as becomes obvious from 
Fig.~\ref{fig:het-mag-dos-OvacTi}. In contrast, the Ti $ 3d $ $ e_g $ 
states, which are located at higher energies as expected from the 
octahedral arrangement of the O atoms about the Ti sites, experience 
a much smaller splitting. The total magnetic moment amounts to 
0.50\,$\mu_{B}$ per full $2\times2$ cell, whereas the magnetic moment 
per single interface plane of 0.22\,$\mu_{B}$ underlines the confinement 
of the magnetic moment in this plane. Qualitatively, our results agree 
well with those of Refs.~\onlinecite{pavlenko2012a,pavlenko2012b}, which, 
however, were obtained for heterostructures with four LAO overlayers 
rather than the three-overlayer heterostructures studied in the present 
work. 

\begin{figure}[tb]
\centering
\includegraphics[width=0.65\linewidth]{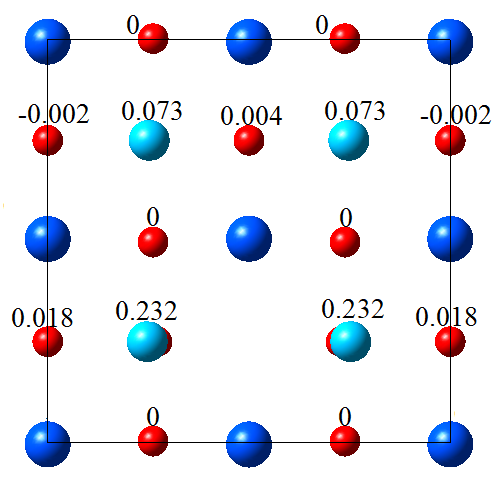}
\caption{Local moments of a 2$\times$2  3\,LAO/4.5\,STO/3\,LAO heterostructure 
         with one oxygen vacancy located at each of the interfaces. Strontium, 
         titanium, and oxygen atoms are given in dark blue, light blue, and 
         red, respectively. }
\label{fig:local_moments_Ovac}
\end{figure}

\begin{figure}[tb]
\centering
\includegraphics[width=0.75\linewidth]{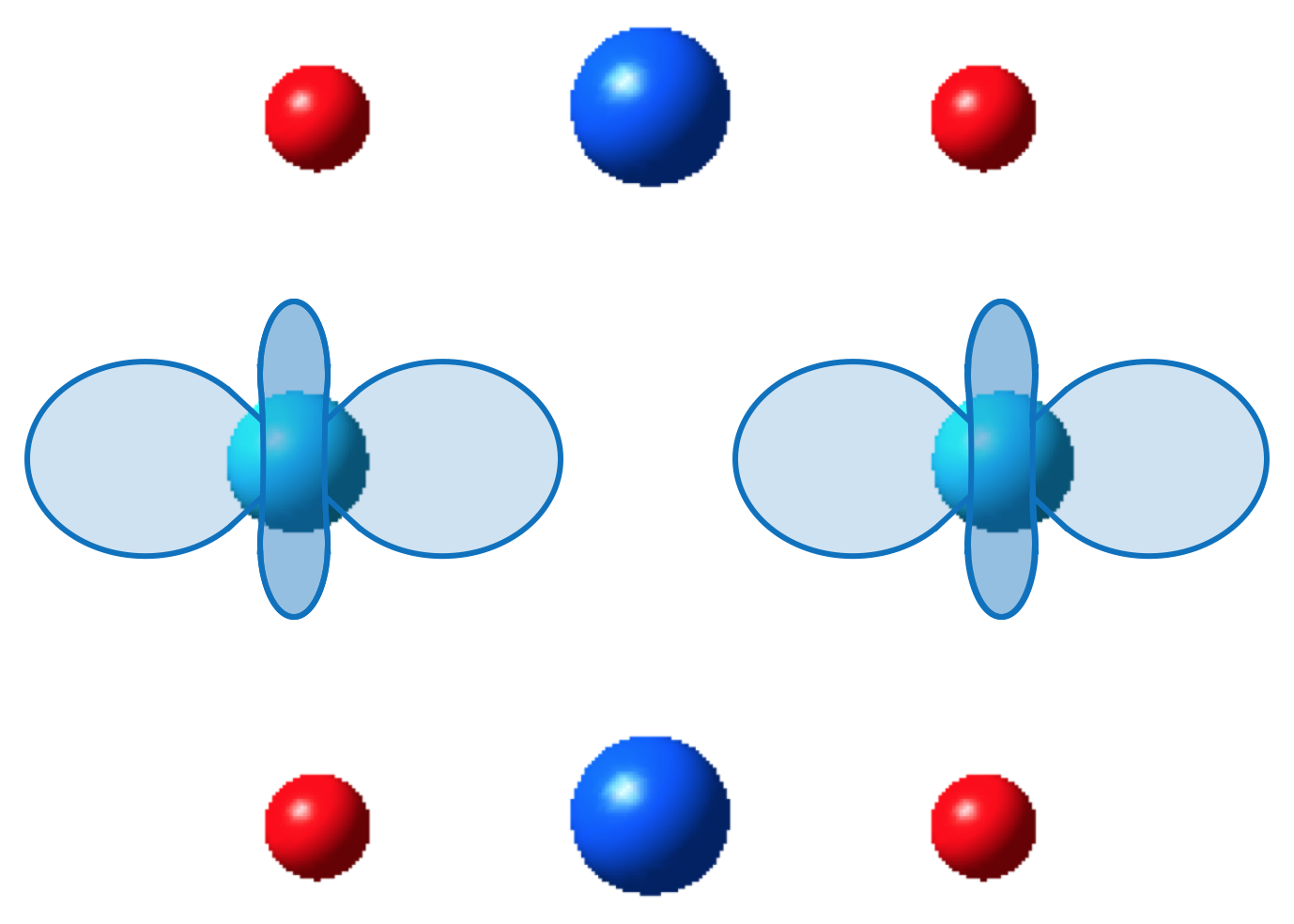}
\caption{Local cluster with an oxygen vacancy and two adjacent Ti atoms,  
         cutout from Fig.~\protect\ref{fig:local_moments_Ovac}. 
         In this cluster the Ti $ 3d e_g $ states with orbital lobes in 
         the direction of the vacancy position are shifted below the 
         Ti $ 3d t_{2g} $ states ("orbital reconstruction" 
         [\onlinecite{pavlenko2012b}]). The standard basis orbitals of 
         the $ e_g $ states, i.e.\ the $ d_{x^2-y^2} $ and $ d_{3z^2-r^2} $ 
         states, are rotated to produce $ d_{y^2-z^2} $ and $ d_{3x^2-r^2} $ 
         states. The $ d_{3x^2-r^2} $ orbitals, which have the lower energy 
         and lobes in the direction of the vacancy, are delineated by 
         a contour line. These orbitals carry a triplet state of the 
         two charge-compensating electrons.}
\label{fig:local_moments_Ovac_lobes}
\end{figure}

The situation changes, when the vacancies are located in the interface 
$ {\rm TiO_2} $ layer. The resulting Ti $ 3d $ partial densities of 
states are displayed in Figs.~\ref{fig:het-mag-dos-Ofar} and 
\ref{fig:het-mag-dos-Onear}, where we distinguish contributions from 
a distant Ti atom and a Ti atom next to the vacancy, respectively. 
Most striking is the downshift of the $ e_g $ states of the latter 
atom as compared to those of the former by about 2~eV, which leads 
to a finite occupation of these orbitals as well as a finite contribution 
to the local magnetic moment, which even exceeds that of the $ t_{2g} $ 
states. In contrast, the $ t_{2g} $ partial densities of states of both 
atoms are very similar. The calculated local magnetic moments are given 
in Fig.~\ref{fig:local_moments_Ovac}. 
They give rise to a magnetic moment of 0.65\,$\mu_{B}$ per single 
interface plane. In contrast, the adjacent layers carry negligible 
magnetic moments of 0.05\,$\mu_{B}$ within the $ 2 \times 2 $ interface 
LaO layer, and 0.02\,$\mu_{B}$ and 0.03\,$\mu_{B}$ within the nearest 
SrO  and TiO$_2$ layers, respectively. The total magnetic moment per 
$ 2 \times 2 $ cell amounts to 1.54\,$\mu_{B}$. Hence, as for the case 
of a surface oxygen vacancy about 85\% of the total magnetization is 
due to the layer comprising the vacancy. 
Moreover, vacancies  in the $ {\rm TiO_2} $ plane neighboring the interface 
plane form a magnetic moment of only 0.16\,$\mu_B$.\cite{pavlenko2012b} 
In the next-nearest neighboring $ {\rm TiO_2} $ plane a magnetic moment 
of approximately 0.08\,$\mu_B$ is identified and for vacancies in the 
third plane from the interface the magnetic moment is approximately zero. 
We thus arrive at the conclusion 
that oxygen vacancies at the interface induce atomically thin magnetic 
layers with a rather uniform background of magnetic moments generated by 
the Ti $ 3d $ $ t_{2g} $ states, which are complemented by well localized 
magnetic moments due to $ e_g $ states of the Ti atoms neighboring the 
vacancy. Of course, as is clear from the local geometry, the latter 
contributions can be even assigned to a single $ e_g $ orbital, namely, 
the one with its lobes pointing towards the vacancy 
(see Fig.~\ref{fig:local_moments_Ovac_lobes}). 
However, in the 
standard representation with the $ 3d_{x^2-y^2} $ and $ 3d_{3z^2-r^2} $ 
orbitals having their lobes parallel and perpendicular, respectively, 
to the interface plane, this is not obvious from the partial densities 
of states. Nevertheless, the situation is not unlike that in $ {\rm V_2O_5} $, 
which has a layered structure with $ {\rm VO_5} $ square pyramids rather 
than $ {\rm VO_6} $ octahedra as basic structural units leading to strong 
downshift of the V $ 3d_{xy} $ states due to reduced bonding-antibonding 
splitting and the formation of the characteristic split-off conduction 
band below about 1~eV the bottom of the other V $ 3d $ bands. \cite{eyert1998}
As before, we point to the good qualitative agreement of our results with 
the previous $+U$ and +DMFT calculations, even though the latter, while 
accounting for quantum fluctuations, lead to considerably reduced magnetic 
moments. \cite{pavlenko2012a,pavlenko2012b,lechermann2014}

\subsection{Hydrogen dopants}
\label{mag-hydrogen}

\begin{figure}[t]
        \centering
        \includegraphics[width=0.65\linewidth]{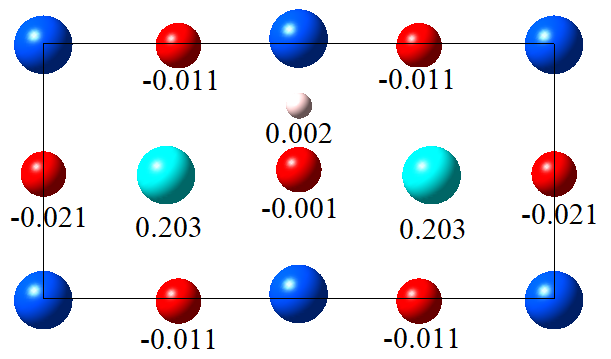}
        \caption{Local moments of a $ 2 \times 1 $ 3\,LAO/4.5\,STO/3\,LAO 
                 heterostructure with a H dopant atom located at each of 
                 the interfaces. Strontium, titanium, oxygen and hydrogen 
                 atoms are given in dark blue, light blue, red, and light 
                 grey, respectively. }
        \label{fig:local_moments_H}
\end{figure}

Finally, spin-polarized calculations were also performed for a 
$ 2 \times 1 $ supercell of the 3\,LAO/4.5\,STO/3\,LAO heterostructure 
with one hydrogen dopant located either in the $ {\rm AlO_2} $ surface 
layer or in the $ {\rm TiO_2} $ interface layer. Whereas the former 
case leads to a negligible magnetization, hydrogen dopants in the 
interface layer induce sizable magnetic moments. The calculated local 
magnetic moments are given in Fig.~\ref{fig:local_moments_H}. 
Specifically, the magnetic moment per 
single $ {\rm TiO_2} $ layer amounts to 0.38\,$\mu_{B}$ (per $ 2 \times 1 $ 
cell) giving rise to a total magnetic moment of 0.82\,$\mu_{B}$.  Hence, 
the confinement of the magnetization to the layer holding the vacancy 
is even stronger than in the previous cases. 

\begin{figure}[b]
\centering
\includegraphics[width=0.9\linewidth]{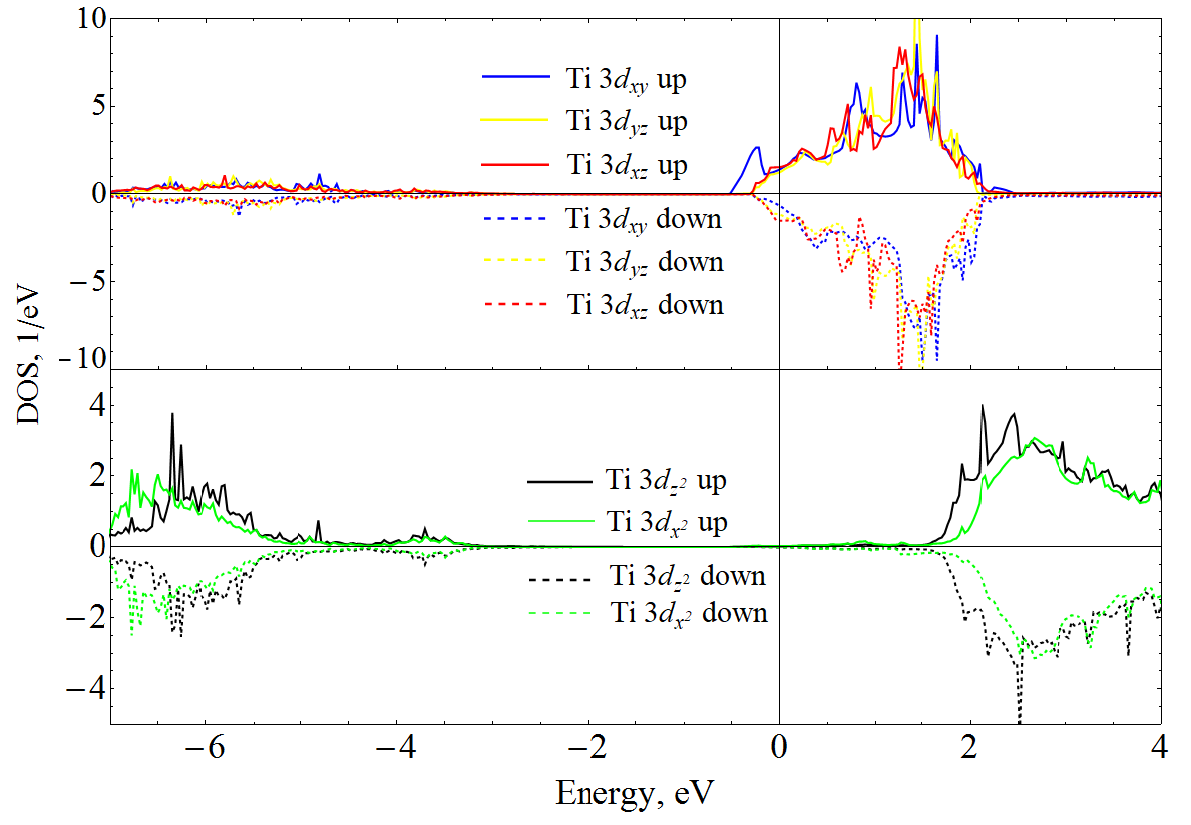}
\caption{Spin-resolved Ti $ 3d $ partial densities of states for 
         a $ 2 \times 1 $ supercell of the 3\,LAO/4.5\,STO/3\,LAO 
         heterostructure with one hydrogen dopant atom located at 
         the interface.}
\label{fig:het-mag-dos-Had}
\end{figure}

The spin-resolved Ti $ 3d $ partial densities of states are shown 
in Fig.\,\ref{fig:het-mag-dos-Had}. 
Obviously, contrasting the situation of an oxygen vacancy in the 
interface layer discussed in the previous section, no downshift of the 
Ti $ 3d $ $ e_g $ orbitals is observed since the octahedral environment 
of the metal atoms is not affected. Moreover, since, as shown in 
Fig.~\ref{fig:local_moments_H}, the hydrogen dopant forms an almost 
rectangular triangle with the neighbouring Ti atoms in the same 
plane, its $ s $-orbital hybridizes mainly with the Ti $d_{xy}$ 
orbitals but much less with the $d_{xz}$ and $d_{yz}$ states. As a 
consequence, electron transfer from the H $ 1s $ orbital is 
predominantly to the $ d_{xy} $ orbitals of the $ 3d $ $ t_{2g} $ 
manifolds centered at the neighboring Ti ions, which thus hold 
the large part of the local magnetic moment. In contrast, the 
$d_{xz}$ and $d_{yz}$ bands contribute only little to the magnetic 
moment as their occupation mainly results from overlap with adjacent 
layers of the interface.  

Finally, we investigated the case of a hydrogen dopant atom filling an 
oxygen vacancy, which, however, again comes with a negligible magnetization.

\section{Summary}
\label{summary}

In the present work, first principles electronic structure calculations 
as based on density functional theory and including local electronic 
correlations within the GGA$ +U $ approach have been employed to study 
the impact of oxygen vacancies and hydrogen dopant atoms on the electronic 
properties of slabs formed from $ {\rm LaAlO_3} $ (LAO) and $ {\rm SrTiO_3} $ 
(STO) as well as of LAO/STO/LAO heterostructure slabs. In particular, 
focus was on the insulating heterostructures with three $ {\rm LaAlO_3} $ 
overlayers. Both kinds of defects cause effective electron-type doping, 
which leads to metallic conductivity in the $ {\rm SrTiO_3} $ slab and 
the $ {\rm LaAlO_3} $/$ {\rm SrTiO_3} $ heterostructure, whereas the 
conductivity of $ {\rm LaAlO_3} $ slabs can be suppressed at suitable doping. 

Calculated energies of defect formation show that hydrogen dopant atoms 
are more favorable at the surface than inside the slab for all structures 
and concentrations considered. The lowest formation energy was found for 
hydrogen adatoms at small concentrations (one hydrogen adatom per 
$ 2 \times 2 $ supercell) due to the reduced Coulomb repulsion. However, 
note that in the present context we considered only the thermodynamic 
stability of defects, which still could be kinetically frozen even in 
case of positive defect formation energies. 

Most strikingly, we find strong decrease of the formation energy of 
hydrogen adatoms on the surfaces of LAO/STO heterostructures as the 
thickness of the LAO overlayer increases. In particular, this energy 
changes sign and becomes negative between two and three LAO overlayers 
for a defect concentration of two hydrogen adatoms per $ 2 \times 2 $ 
surface cell. 
Taken together with the metallic conductivity induced 
by hydrogen dopant atoms, these findings would provide a consistent explanation for 
the semiconductor-metal transition observed for LAO/STO heterostructures 
at increasing thickness of the overlayers. The polar catastrophe mechanism 
without dopants has several drawbacks, as presented in Sec.~\ref{intro}.
Therefore, an alternative mechanism is of relevance, especially since 
the binding of H adatoms at the surface is thickness dependent and the critical thickness is a few LAO 
layers. Finally, the surface 
passivation scenario also explains the suppression of the shift of the 
effective potential across the LAO overlayer.

Contrasting the preference of hydrogen atoms for surface sites, the 
formation energies of oxygen vacancies develop local minima at the 
interface of LAO/STO heterostructures for high enough defect 
concentrations. Motivated by this finding as well as by previous 
reports that oxygen vacancies cause magnetism
we confirmed the local-moment formation induced by vacancies 
located either at the surface or in the $ {\rm TiO_2} $ interface 
layer and found strong confinement of the magnetization within these  
layers. In addition, vacancies in the $ {\rm TiO_2} $ interface layer 
lead to drastic energetical downshift of the $ 3d $ $ e_g $ orbitals 
of the Ti atoms neighboring the vacancy, which carry large part of the 
well localized magnetic moment. Magnetism in the interface layer 
thus emerges from two distinct sources: While the Ti $ 3d $ $ t_{2g} $ 
orbitals form a rather uniform atomically thin background magnetization, 
the $ e_g $ orbitals of the Ti atoms neighboring the vacancy add 
point-like localized magnetic moments. 

Finally, hydrogen dopant atoms in the $ {\rm TiO_2} $ interface layer 
of the LAO/STO heterostructure give rise to magnetism with even 
larger magnetic moments, which are mainly carried by the Ti $ 3d_{xy} $ 
states and again confined within the interface layer.

\section*{Acknowledgments}

We thank Dirk Fuchs, Jochen Mannhart, and Michael Sing for stimulating 
discussions.
The research was carried out using the equipment of the shared research 
facilities of HPC computing resources at Lomonosov Moscow State University. 
The authors from KFU acknowledge partial support by the Program of 
Competitive Growth of Kazan Federal University. 
The work of I.~I.~Piyanzina was funded by a subsidy allocated to Kazan 
Federal University for the state assignment in the sphere of scientific 
activities (project 3.9779.2017/8.9) and by the German Science Foundation 
(DFG) -- grant number 107745057 -- TRR\,80. All authors are grateful for 
the support by the TRR\,80. T.~Kopp wants to express his gratitude to the 
late Natalia Pavlenko who understood so much of this physics and who was 
a brilliant coworker.

\end{document}